\documentclass[hidelinks,11pt]{article}
\usepackage{amsmath}
\usepackage{amsfonts}
\usepackage{color,soul}
\usepackage[table]{xcolor}
\usepackage{natbib}
\usepackage{threeparttable} 
\usepackage{hyperref}
\usepackage{booktabs}
\usepackage{graphicx}
\usepackage{epstopdf}
\usepackage{lipsum}
\usepackage{cleveref}
\usepackage{rotating}
\usepackage{caption}
\usepackage{subcaption}
\usepackage{multirow}
\usepackage[stable]{footmisc}
\usepackage{breqn}
\usepackage{booktabs}
\usepackage{rotating}
\usepackage{floatrow}
\usepackage{mathtools}
\usepackage{breqn}

\usepackage{array}
\makeatletter
\g@addto@macro{\endtabular}{\rowfont{}}
\makeatother
\newcommand{\rowfonttype}{}
\newcommand{\rowfont}[1]{
   \gdef\rowfonttype{#1}#1%
}
\newcolumntype{L}{>{\rowfonttype}l}
\newcolumntype{C}{>{\rowfonttype}c}

\title{Panel quantile regressions for estimating and predicting the Value--at--Risk of commodities
}

\author{Franti\v{s}ek \v{C}ech\thanks{Institute of Economic Studies, Charles University, Opletalova 26, 110 00, Prague,  CR and Institute of Information Theory and Automation, Academy of Sciences of the Czech Republic, Pod Vodarenskou Vezi 4, 182 00, Prague, Czech Republic.Phone: +420 776 535 106 Email: frantisek.cech@fsv.cuni.cz} \and Jozef Barun\'{\i}k\thanks{Institute of Economic Studies, Charles University, Opletalova 26, 110 00, Prague,  CR and Institute of Information Theory and Automation, Academy of Sciences of the Czech Republic, Pod Vodarenskou Vezi 4, 182 00, Prague, Czech Republic. Phone: +420 776 259 273. Email: barunik@utia.cas.cz}} 

\begin{document}
\maketitle

\begin{abstract}
\noindent 
This paper investigates how realized and option implied volatilities are related to the future quantiles of commodity returns. Whereas realized volatility measures ex-post uncertainty, volatility implied by option prices reveals the market's expectation and is often used as an ex-ante measure of the investor sentiment. Using a flexible panel quantile regression framework, we show how the future conditional quantiles of commodities returns depend on both ex-post and ex-ante uncertainty measures. Empirical analysis of the most liquid commodities covering main sectors including energy, food, agricultural, precious and industrial metals reveal several important stylized facts about the data. We document common patterns of the dependence between future quantile returns and ex-post as well as ex-ante volatilities. We further show that conditional returns distribution is platykurtic and time-invariant. The approach can serve as a useful risk management tools for investors interested in commodity future contracts.

\end{abstract}

\newpage
\section{Introduction}

Commodities play an increasingly significant role in the asset allocations of institutional investors, and with the onset of exchange-traded funds became a regular asset class. Academic debate spurred by the developments provide valuable insights into the economics of commodity markets, as well as to several crucial aspects as price forecasting, risk measurement or hedging. One of the main challenges faced by researchers is the fact that commodities are non-homogeneous assets, and their risk and also return, may differ substantially since each commodity is driven by specific supply and demand forces. A traditional economist's view on an asset price being a stream of future discounted expected cash flows is hence not directly applicable, and pricing of commodities is instead driven by short-term variations in the supply. In addition, exogenous factors like weather conditions, inventory levels, storage costs, production shocks, or even geopolitical events play a crucial role rendering risk measurement a difficult task. In this paper, we propose a simple, robust framework that can be used to model and forecast Value--at--Risk (VaR) of commodities semi-parametrically without the need of traditional assumptions. Empirical results support our approach, and we uncover stylized facts useful for the investors as well as policymakers.  

Complex nature of commodity pricing results in risk characteristics that are different from those in financial assets such as stocks, bond, and currencies. Return distributions, as measured by volatility, skewness, kurtosis, and empirical quantiles are different from traditional asset classes; hence we need more flexible techniques to measure the risks. Many researchers have tried to model the Value--at--Risk of commodities without reaching consensus about the appropriate model. The three main approaches used in the literature are useful but lack the ability to work with complexness of the commodity data. First, RiskMetrics \citep{Longerstaey1996} does not necessarily capture the correct return distribution conditional on the changing volatility. Second, the historical simulation used for example by \cite{DAVIDCABEDO2003239} has an opposite problem that it captures empirical return distribution, but does not make it conditional on volatility. Third, more advanced parametric models mostly built within the family of Generalized AutoRegressive Conditional Heteroskedasticity (GARCH) models improve fits \citep{GIOT2004379,ALOUI20102326,youssef2015value,LUX2016117,HUNG20081173,CHIU2010423}, however, require fat-tailed distributions, long memory, as well as other features that lead to heavy parametrization making the approach less tractable.

Since the seminal work of \cite{koenker1978regression}, quantile regression models have been increasingly used in many disciplines. Notable contributions in finance are by \cite{engle2004caviar} who were among the first to use quantile regression and develop the Conditional Autoregressive Value--at--Risk (CAViaR) model. Importantly to our work, \cite{Zikes_Barunik2015semi} show that various realized measures are useful in forecasting quantiles of future returns without making assumptions about underlying conditional distributions. The resulting semi-parametric model captures conditional quantiles of financial returns well in a flexible framework. Moving the research focus towards the multivariate framework, and concentrating on interrelations between quantiles of more assets, \cite{white2015var} pioneers the extension. The different stream of multivariate quantile regression based literature concentrates on the analysis using factors \citep{chen2016quantile,ando2017quantile}, but the research is recent and awaits its development. Although the application of quantile regression to forecasting quantiles of various economic variables is not new in finance, it has rarely been applied in the context of commodities. Among few, \cite{LI201760} adapts quantile regression for forecasting day-ahead electricity load quantiles, and \cite{REBOREDO201633} studies quantile dependence of oil price movements and stock returns.

With this respect, work by \cite{Zikes_Barunik2015semi} is important as it provides a link between future quantiles of return distribution and its past variation. Despite the non-homogeneous nature of the commodities, \cite{christoffersen2018factor} uncovered several stylized facts pointing to factor structure in volatility. Being interested in future quantiles of the commodity returns distributions, it is tempting to ask, if there is a common structure in the quantiles of commodity returns. Inspired by our previous findings on financial markets \citep{cech2017measurement}, we hypothesize there might be useful commonalities to be uncovered. In the quantile regression set-up, no similar study uncovers information captured in the panels of volatility series of commodity markets. Hence our work can possibly open new questions in modeling VaR of commodity markets. 

This paper contributes to literature by identifying common patterns of the dependence between future quantiles of commodity returns and ex-post/ex-ante volatility measures using flexible panel quantile regression approach. Our simple, yet robust modelling strategy utilize all the advantages offered by panel quantile regression and commodity datasets. We document interesting empirical regularities by controlling for otherwise unobserved heterogeneity among commodities. In particular, we reveal common factors in volatility that have direct influence on the future quantiles of their returns. Our research is important since current literature knows little about the potential of the uncertainty factors in the precise identification of the extreme tail events of the commodity returns distribution. More importantly, even less is known about commonalities between more commodities with this respect. Our research attempts to contribute in this direction.

In the first part of our empirical application, we study the behavior of energy (Crude Oil, Natural Gas), precious metals (Gold, Silver), industrial metal (Copper), agricultural (Cotton) and food (Corn) commodities during a period of the Global Financial Crisis. We document common effects of the ex-post uncertainty measured by realized volatility on the estimation of the Value--at--Risk of commodities. These effects are stable over time and do not dramatically change when we compare the results from the crisis and after-crisis period. In contrast to our expectations, we document homogeneous behavior across commodities. Moreover, the conditional distribution of the returns standardized by their realized volatility is \textit{platykurtic} as opposed to previous parametric studies where a variety of GARCH models were used. To match the empirical data, GARCH models needed  fat-tailed distribution \citep{giot2003market, marimoutou2009extreme, cheong2009modeling, charles2017forecasting} or combination with Extreme Value Theory \citep{youssef2015value}. Since the commodities are considered to be relatively less risky in comparison to financial assets \citep{bodie1980risk, gorton2006facts, conover2010now} our findings are in line with \cite{andersen2000exchange} who document that returns of financial assets standardized by their realized volatility are almost Gaussian. Our findings can be attributed to flexibility offered by the framework we use. Our model does not need an assumption about the distribution and estimates volatility non-parametrically.

In the second part, we employ option implied volatility as an ex-ante measure of uncertainty and relate it to future returns quantiles. Volatility implied by option prices reveals the market's expectations and is often used as an ex-ante measure of the investor sentiment. Relatively recently, new indices measuring the market's expectation of 30-day volatility of commodity prices by applying the well-known Volatility Index (VIX) methodology have been introduced for more commodities. Still, availability of the commodity option implied volatility indexes is limited, therefore we concentrate on the main ones: the Crude Oil, Gold and Silver. We document patterns driving Value--at--Risks of the selected commodities that are similar for both ex-post and ex-ante volatility measures. Moreover, once we control for ex-post uncertainty, the ex-ante volatility shows the great importance for the Value--at--Risk estimation.

\section{Theoretical Background} \label{sec:theory}
\subsection{Value--at--Risk and modelling quantiles of returns}
Value--at--Risk introduced by J.P.Morgan in 1994 quickly became an industry standard for risk measurement in finance and still attracts great attention of researchers. The popularity of the VaR steams from its simplicity since it represents a maximum potential loss of a portfolio at given probability as a single number. According to \cite{Longerstaey1996} VaR can be parametrically calculated as:
\begin{equation}\label{eq: VaR}
VaR_\tau=\gamma_\tau \sigma,
\end{equation}
where $\gamma_\tau$ is the $\tau$ quantile of the standard normal distribution and $\sigma$ is the volatility of the asset. 

Growing financialization of commodity markets,\footnote{For a detailed overview see \cite{cheng2014financialization}.} motivates researchers to apply standard time series techniques, well established in the financial industry, to study riskiness of the commodities. Many researchers, therefore, study Value--at--Risk using parametric approach, where volatility in \autoref{eq: VaR} comes from a variety of GARCH models. The recent attempts include \cite{youssef2015value} who combine long-memory GARCH models with Extreme Value Theory; \cite{LUX2016117} where  Markov-switching multifractal and various GARCH models are applied to model and forecast oil price volatility; \cite{chkili2014volatility} where wide range of linear and non-linear GARCH models is used to study VaR of the energy and precious metals commodities; or \cite{giot2003market} where it is shown that skewed Student APARCH works best in forecasting the VaR in the commodity markets.

More general definition of Value--at--Risk presented in \cite{jorion2007value} suggest to think about VaR as being the quantile of the projected distribution of returns over the certain period of time. In this spirit, \cite{Zikes_Barunik2015semi} show that Value--at--Risk estimation can be formulated as a quantile regression of returns on their ex-post volatility without making any distributional assumptions. Commonalities in the panels of Realized Volatilities \citep{bollerslev2016risk} motivate \cite{cech2017measurement} to extend previous work of \cite{Zikes_Barunik2015semi} into multivariate space and study common factors in volatility that have a direct
influence on the future quantiles of returns. In particular, they propose to control for otherwise unobserved heterogeneity among financial assets and measure common market risk factors using panel quantile regression approach.

An important link between commodities and stocks provide work of \cite{christoffersen2018factor} who find a strong relationship between commodity and stock market volatility. In our work, we hypothezise that commodities share  similarities with stock market in the behavior of the conditional return distributions. To identify common patterns driving the Value--at--Risk of commodities, we adopt Panel Quantile Regression Model for Returns \citep{cech2017measurement}. In particular, we study a quantile pricing equation of the following form:
\begin{equation}\label{eq:panel quantile dynamics}
Q_{r_{i,t+1}}(\tau\lvert X_{i,t})=\alpha_i(\tau)+X_{i,t}^{\top}\beta(\tau),
\end{equation} 
where $\tau \in (0,1)$; $r_{i,t+1}=p_{i,t+1}-p_{i,t}$ are logarithmic daily returns; $X_{i,t}$ is a matrix of ex-post and/or ex-ante volatility measures; and $\alpha_i$ represents individual fixed effects. 

The model defined in \autoref{eq:panel quantile dynamics} allow us to study the influence of the ex-post/ex-ante uncertainty on the specific quantiles of the commodity returns through the $\beta$ coefficients that are common for all commodities, and account for unobserved heterogeneity among assets represented by individual fixed effects, $\alpha_i$. To obtain parameter estimates, we apply fixed effect estimator of \cite{koenker2004quantile}\footnote{We refrain from using penalization as originally suggested by \cite{koenker2004quantile} - our cross-sectional dimension is much smaller than the time dimension, so the number of estimated parameters is small.} and solve:
\begin{equation} \label{eq:PQR equation}
\min\limits_{\alpha_i(\tau),\beta(\tau)}\sum_{t=1}^{n}\sum_{i=1}^{t_i} \rho_{\tau}\left(r_{i,t+1}-\alpha_i(\tau)-X_{i,t}^{\top}\beta(\tau)\right), 
\end{equation}
where $\rho_{\tau}(u)=u\left(\tau-I(u(<0))\right)$ is the quantile loss function defined in \cite{koenker1978regression}.

\subsection{Measures of uncertainty}
We distinguish between ex-post and ex-ante uncertainty about the commodity price. The former measures variation in the historical data, while the latter describes the market's sentiment and expectation of future risk. Although underlying commodity is the same in both cases, the information content of each measure might differ \citep{giot2007information}.

Ex-post uncertainty is represented by the Realized Volatility estimated at the 5-minutes frequency. The choice of uncertainty measure is motivated by the results of \cite{liu2015does} who show that it is hard to beat the performance of 5-min RV using more sophisticated realized measures. We construct the estimator as a square-root of sum of the squared intraday returns \citep{Andersen2003} 
\begin{equation}
\widehat{RV}^{1/2}_{i,t}=\sqrt{\sum_{k=1}^N \left(\Delta_k p_{i,t} \right)^2},
\end{equation} 
where $\sum_{k=1}^N \left(\Delta_k p_{i,t} \right)^2$ is the Realized Variance estimator with $\Delta_k p_{i,t} = p_{i,t-1+\nu_k/N}-p_{i,t-1+\nu_{k-1}/N}$ being a discretely sampled vector of $k$-th intraday log-returns of $i$th commodity in $[t-1,t]$, with $N$ intraday observations.

\begin{figure}[H]
\centering
\caption{Daily Returns} 
\begin{subfigure}[b]{0.24\textwidth}
\caption{Crude Oil}\label{fig:ret CL}
\vspace{-10pt}
\includegraphics[width=\textwidth, height=0.8\textwidth]{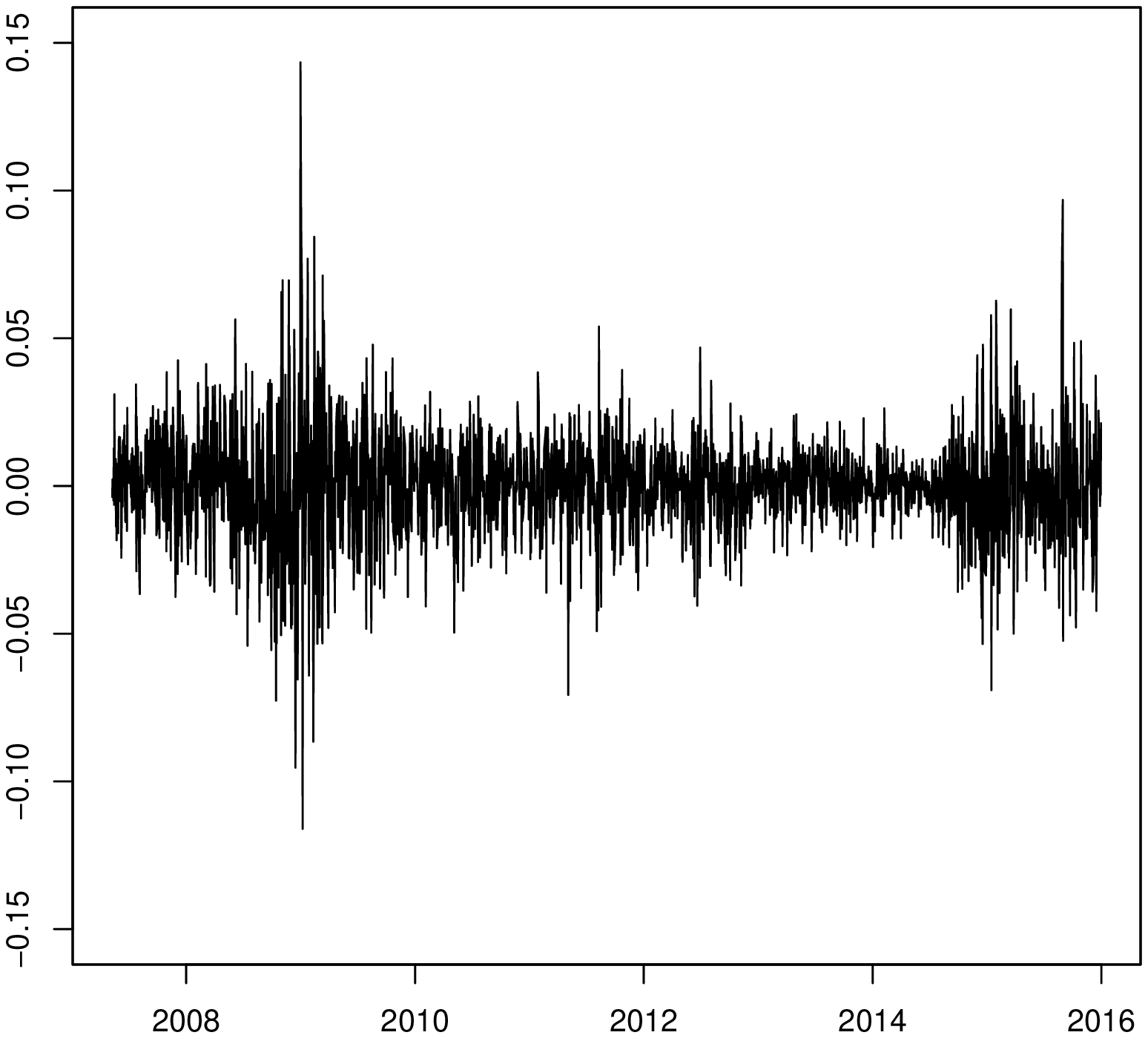}
\end{subfigure}
\begin{subfigure}[b]{0.24\textwidth}
\caption{Corn}\label{fig:ret CN}
\vspace{-10pt}
\includegraphics[width=\textwidth, height=0.8\textwidth]{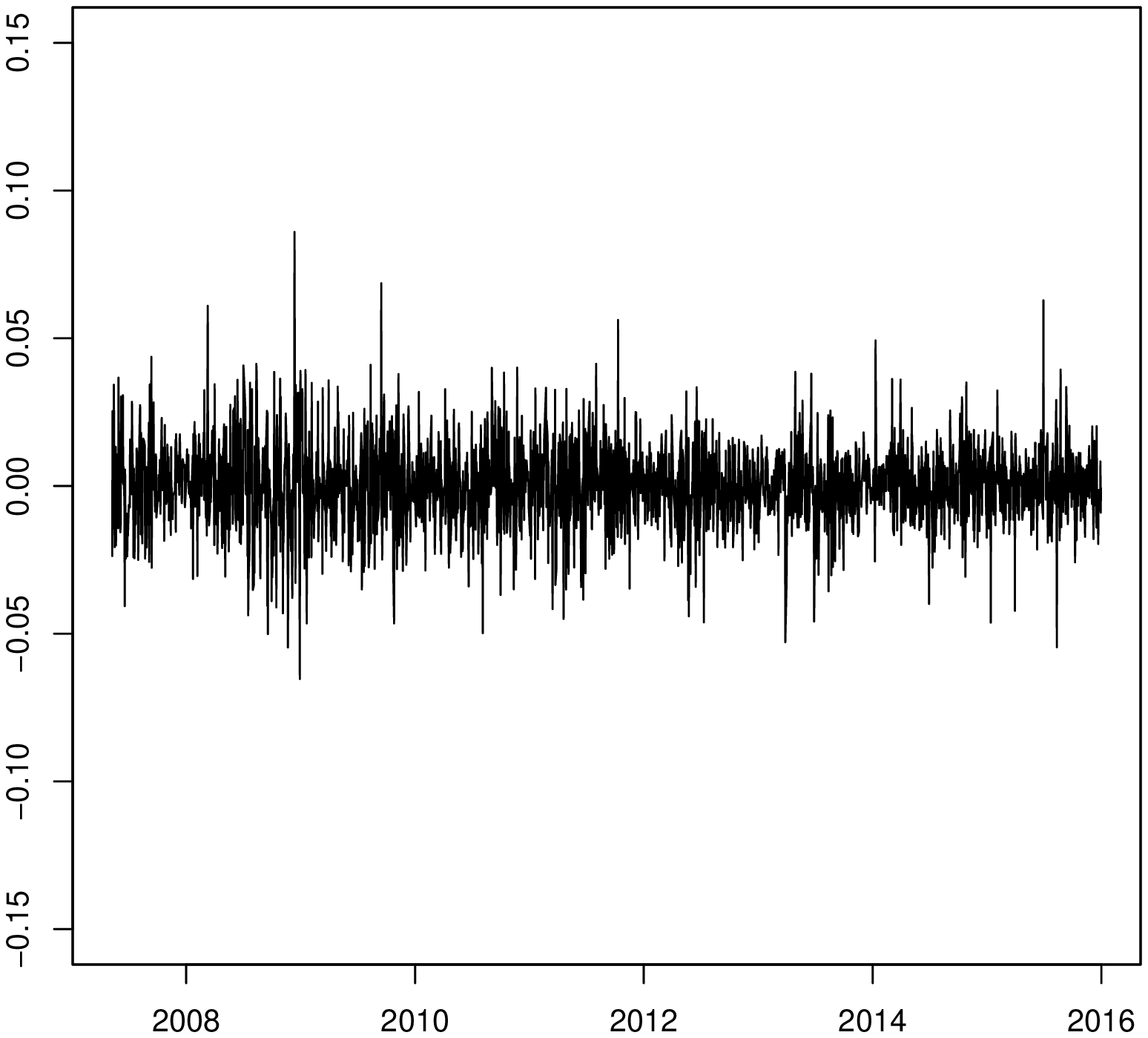}
\end{subfigure}
\begin{subfigure}[b]{0.24\textwidth}
\caption{Cotton}\label{fig:ret CT}
\vspace{-10pt}
\includegraphics[width=\textwidth, height=0.8\textwidth]{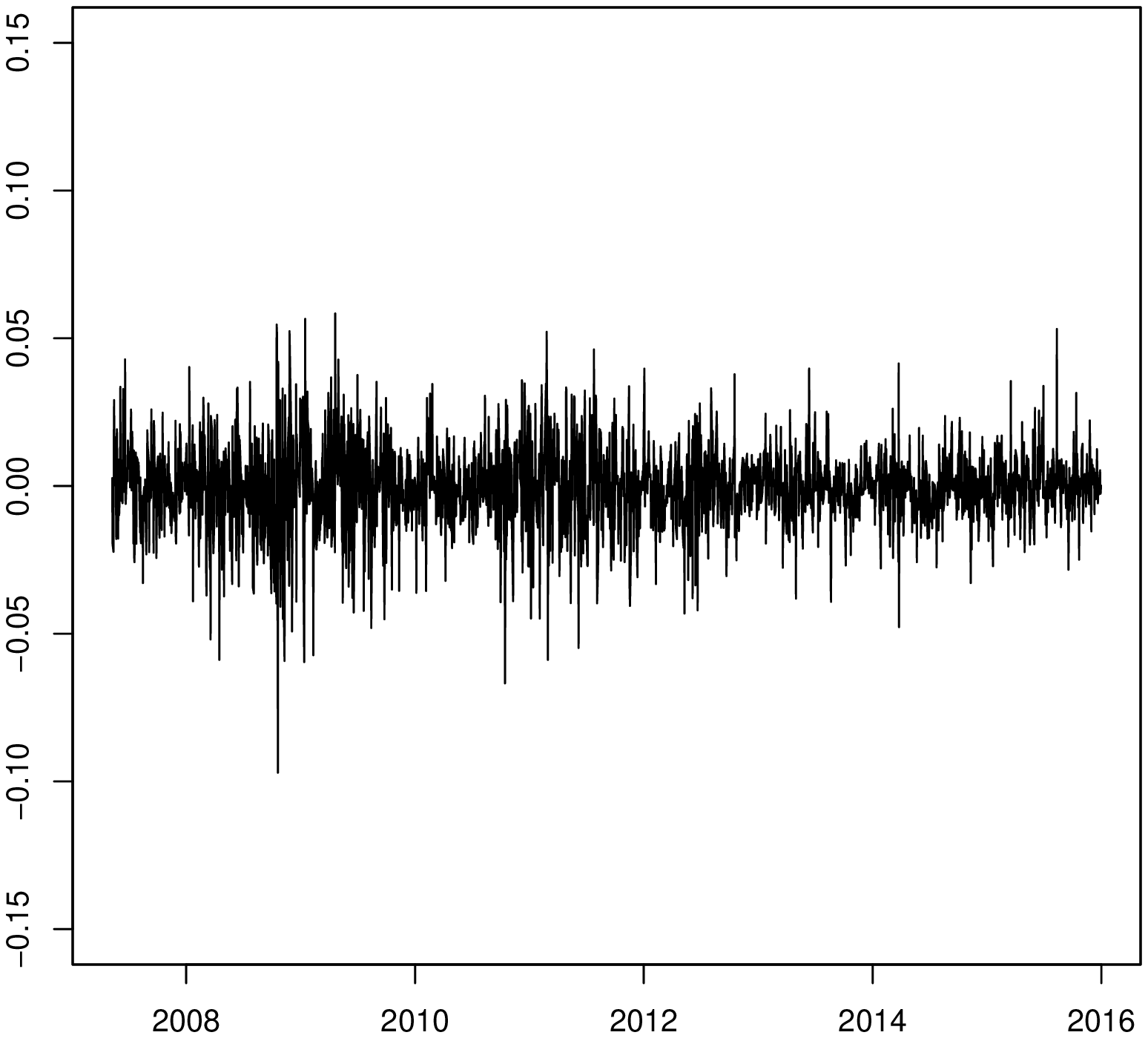}
\end{subfigure}
\begin{subfigure}[b]{0.24\textwidth}
\caption{Gold}\label{fig:ret GC}
\vspace{-10pt}
\includegraphics[width=\textwidth, height=0.8\textwidth]{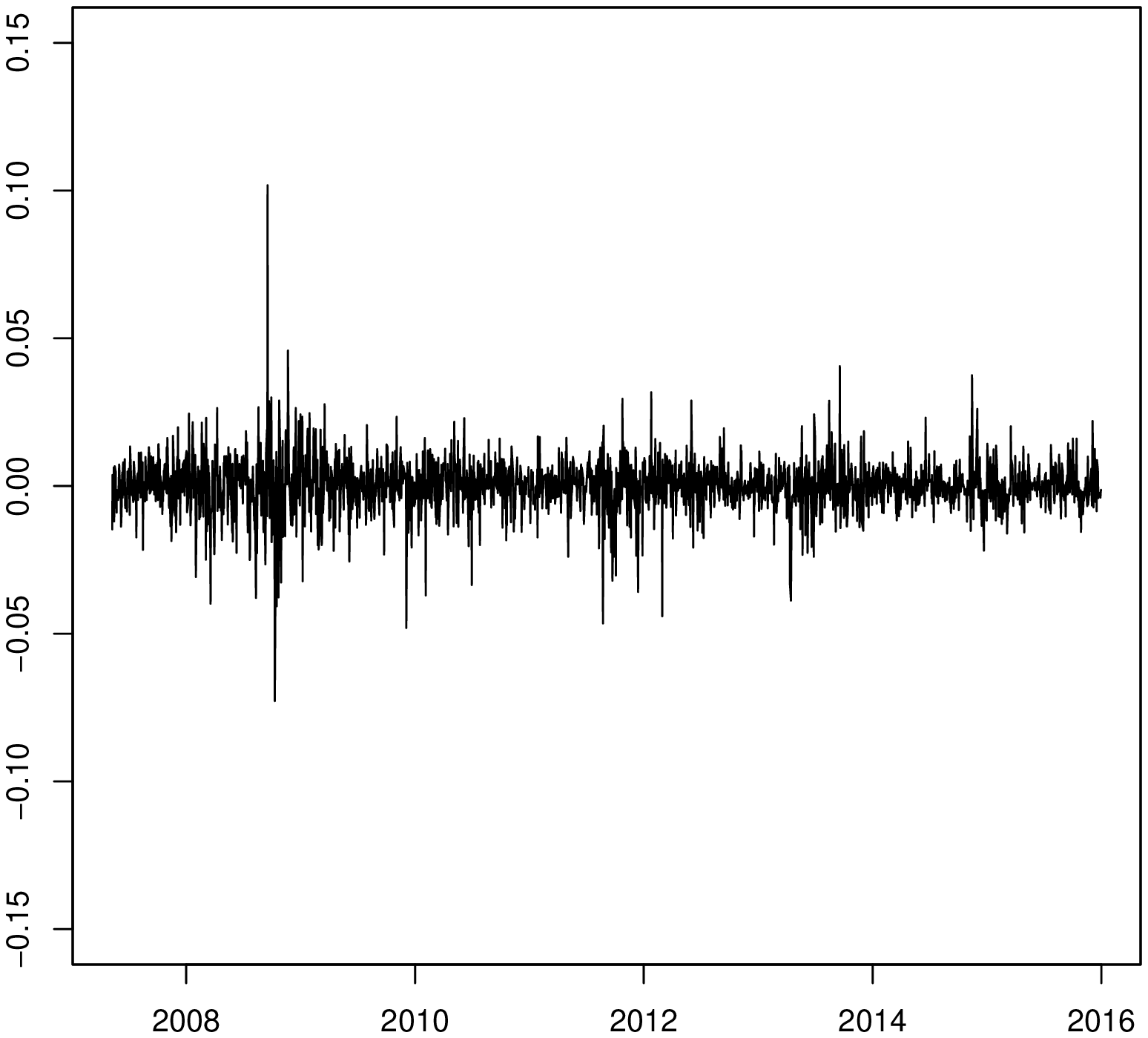}
\end{subfigure}
\begin{subfigure}[b]{0.24\textwidth}
\caption{Copper}\label{fig:ret HG}
\vspace{-10pt}
\includegraphics[width=\textwidth, height=0.8\textwidth]{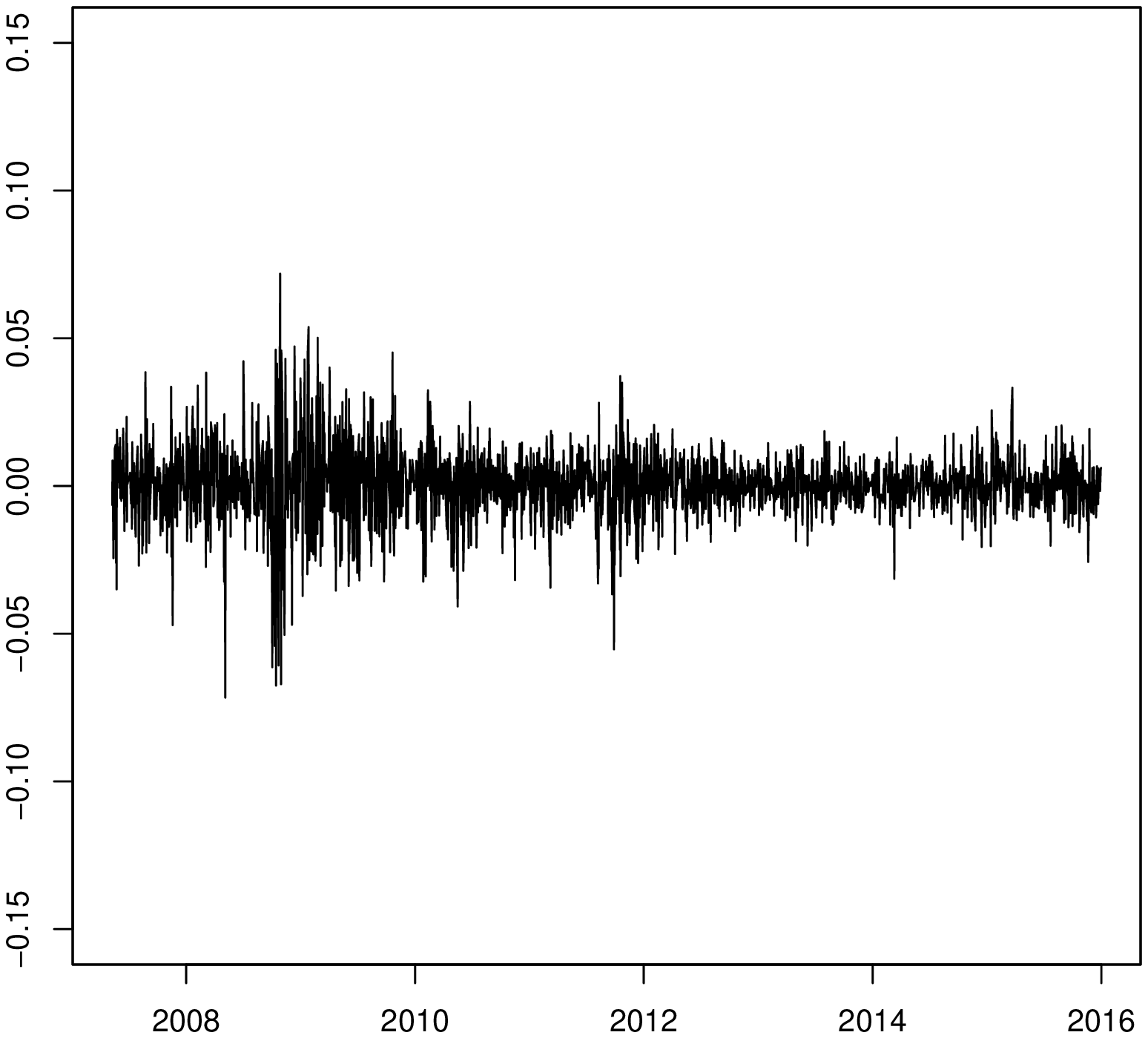}
\end{subfigure}
\begin{subfigure}[b]{0.24\textwidth}
\caption{Natural Gas}\label{fig:ret HG}
\vspace{-10pt}
\includegraphics[width=\textwidth, height=0.8\textwidth]{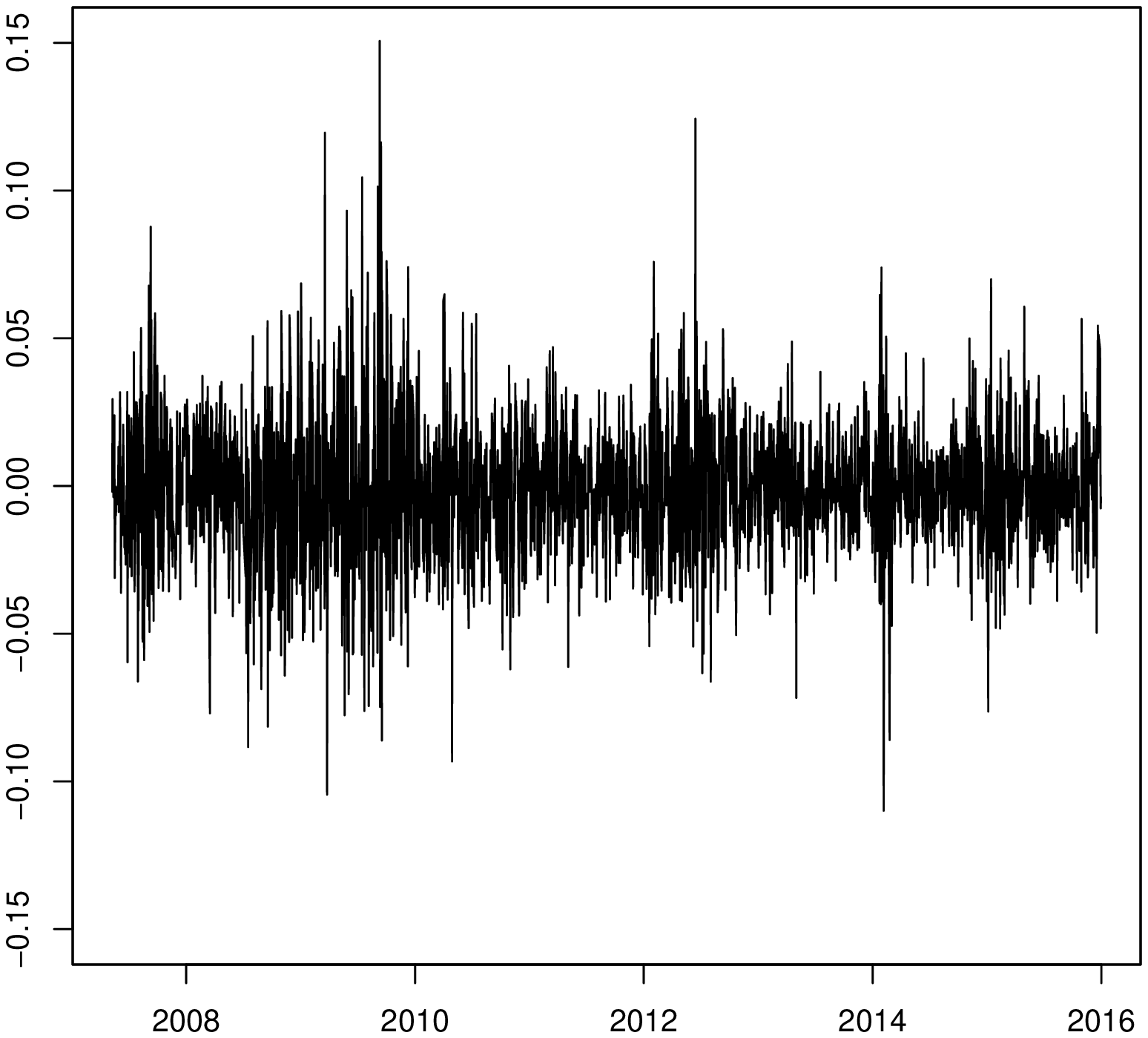}
\end{subfigure}
\begin{subfigure}[b]{0.24\textwidth}
\caption{Silver}\label{fig:ret HG}
\vspace{-10pt}
\includegraphics[width=\textwidth, height=0.8\textwidth]{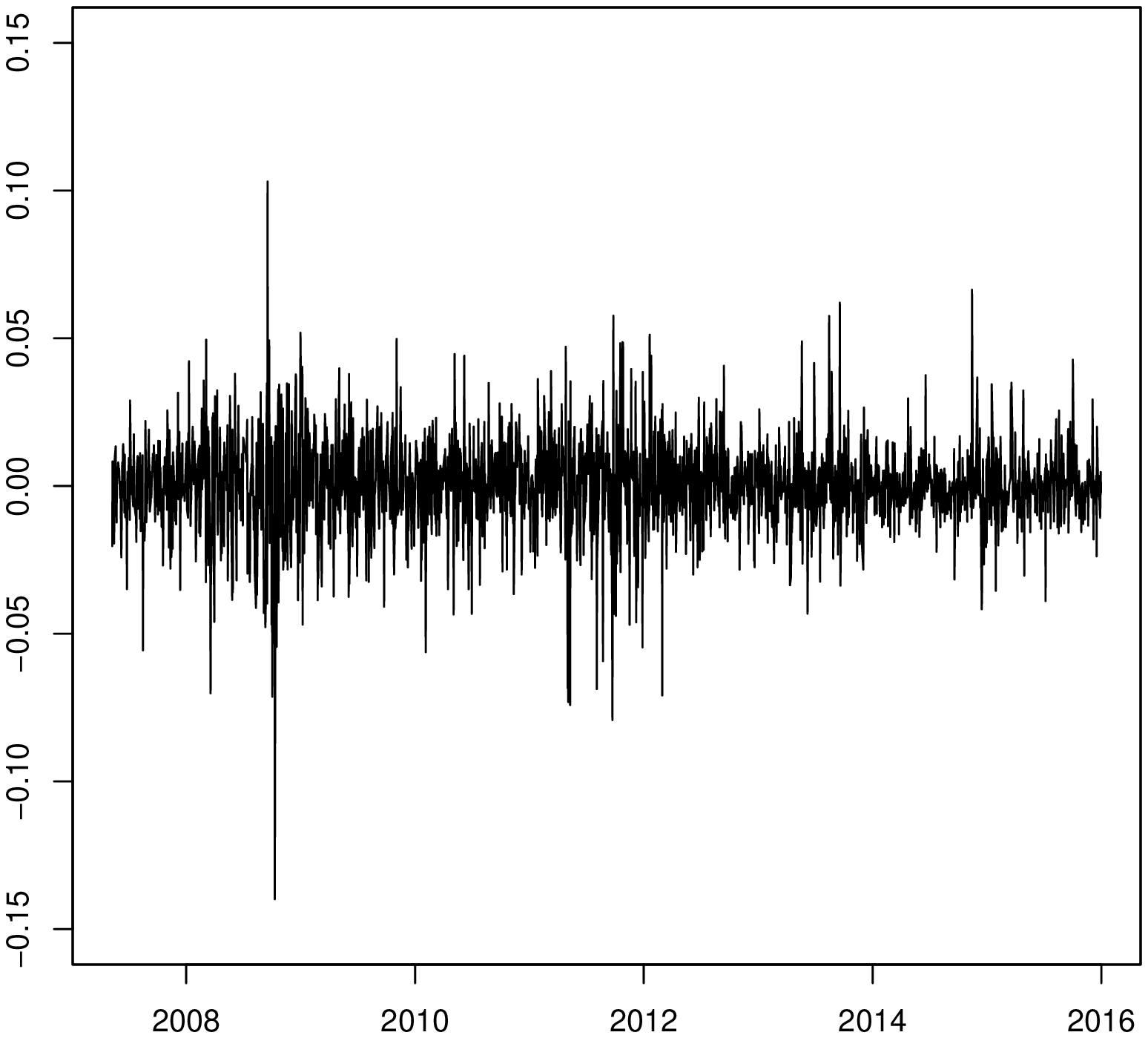}
\end{subfigure}
\captionsetup{justification=centering}
\floatfoot{Note: Plot displays open-close returns during period May 10, 2007 - December 31, 2015.}
\label{fig: returns}
\end{figure}

\begin{figure}[ht]
\centering
\caption{Realized Volatility} 
\begin{subfigure}[b]{0.24\textwidth}
\caption{Crude Oil}\label{fig:RV CL}
\vspace{-10pt}
\includegraphics[width=\textwidth, height=0.8\textwidth]{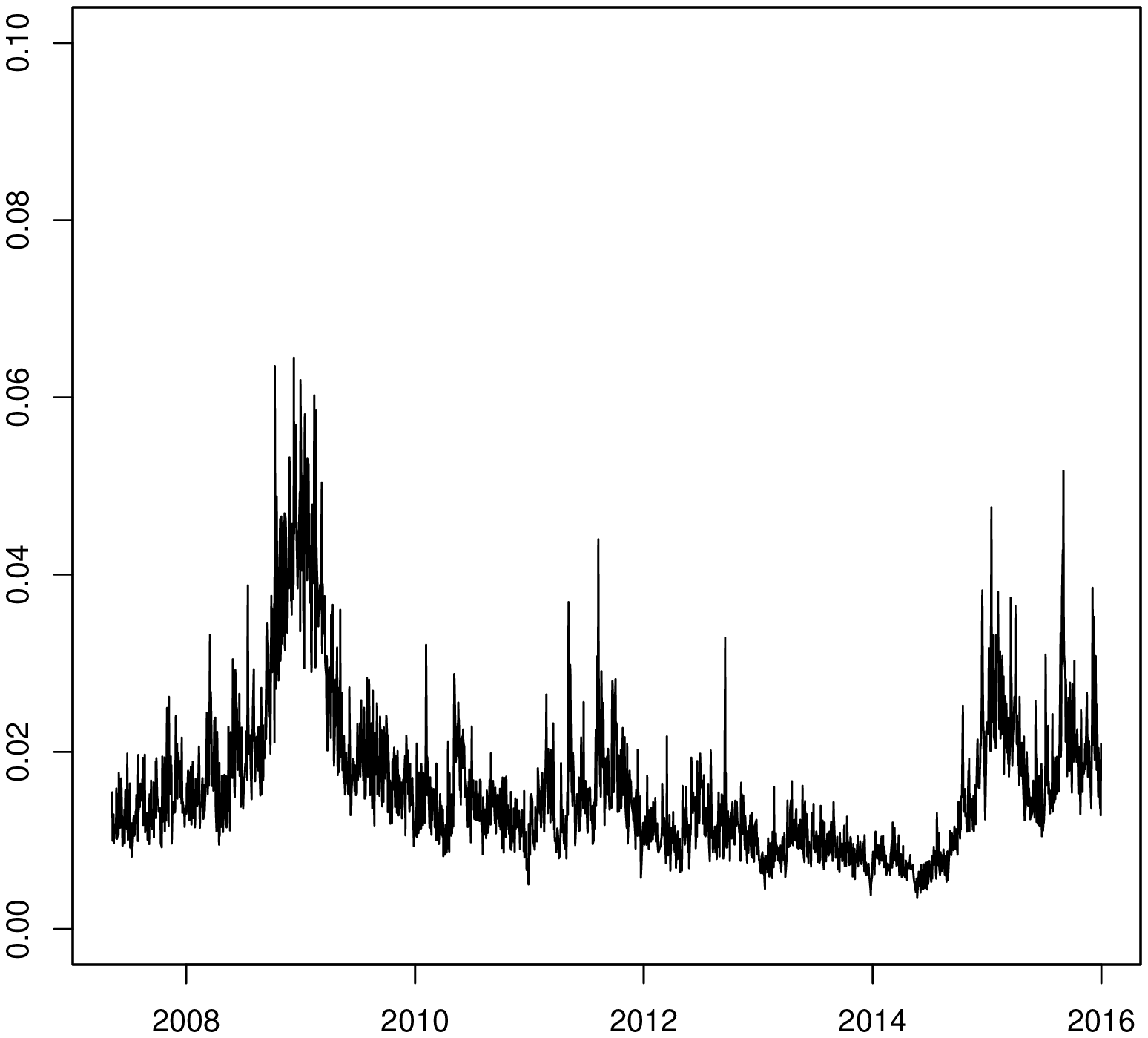}
\end{subfigure}
\begin{subfigure}[b]{0.24\textwidth}
\caption{Corn}\label{fig:RV CN}
\vspace{-10pt}
\includegraphics[width=\textwidth, height=0.8\textwidth]{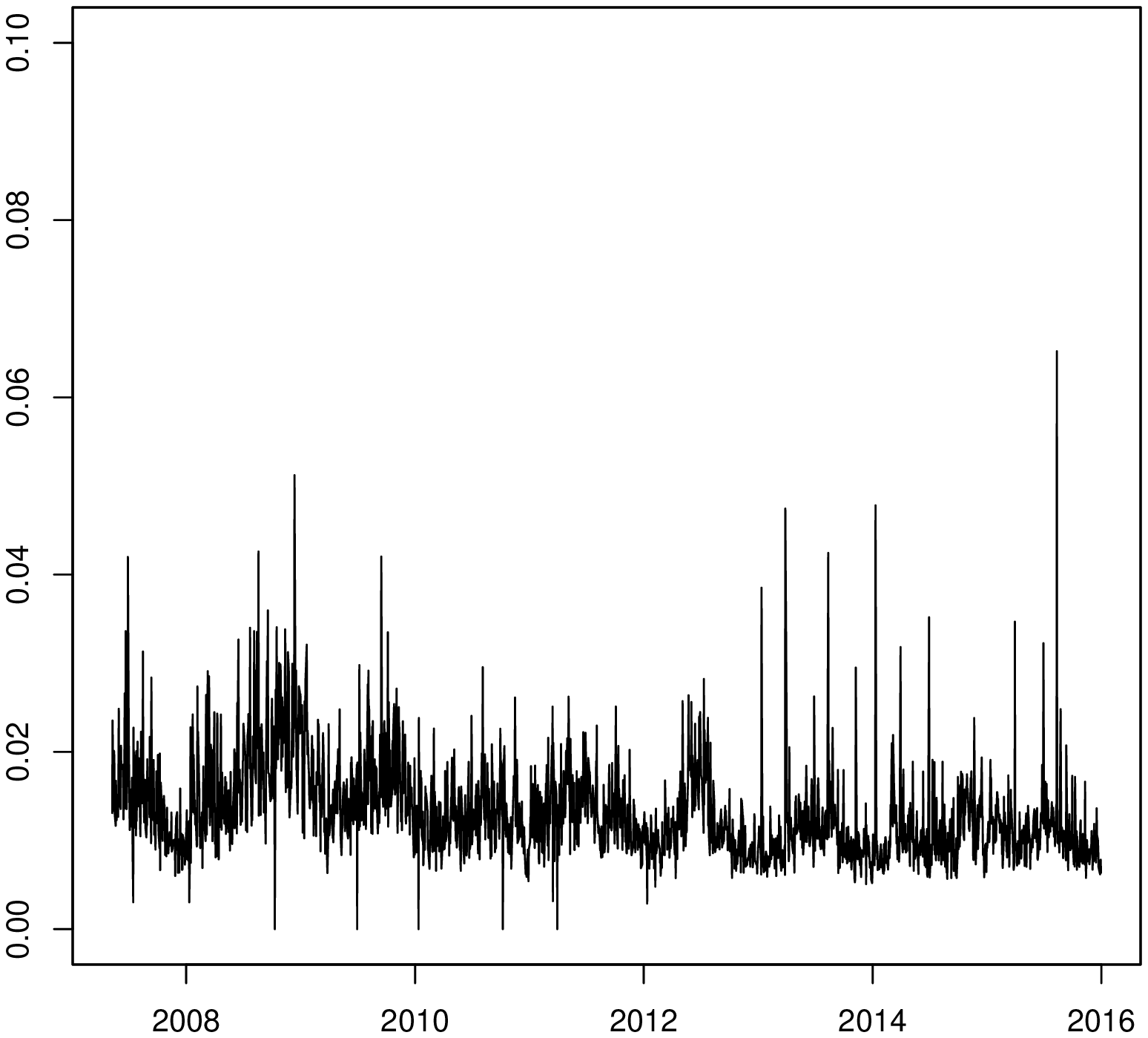}
\end{subfigure}
\begin{subfigure}[b]{0.24\textwidth}
\caption{Cotton}\label{fig:RV CT}
\vspace{-10pt}
\includegraphics[width=\textwidth, height=0.8\textwidth]{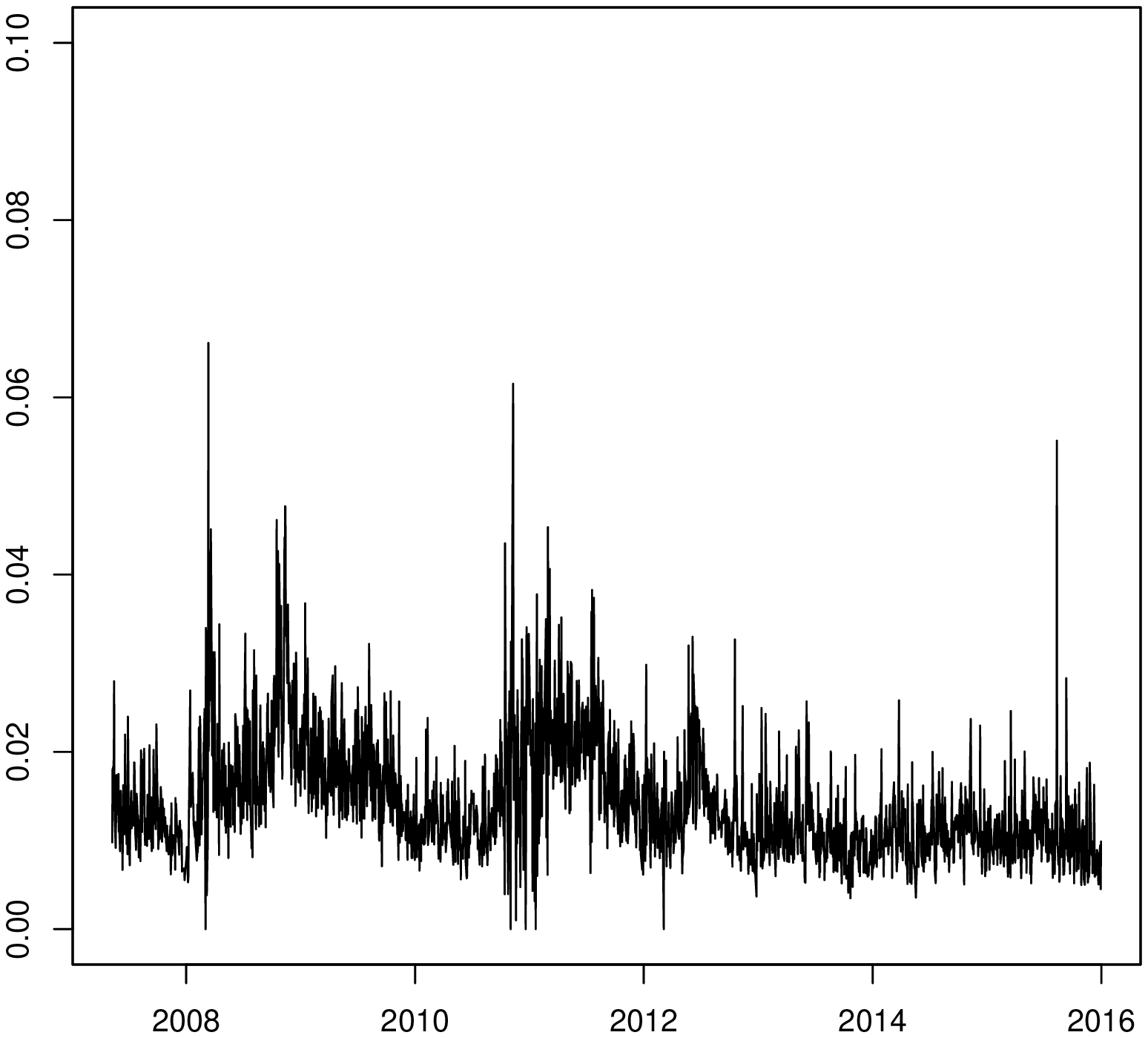}
\end{subfigure}
\begin{subfigure}[b]{0.24\textwidth}
\caption{Gold}\label{fig:RV GC}
\vspace{-10pt}
\includegraphics[width=\textwidth, height=0.8\textwidth]{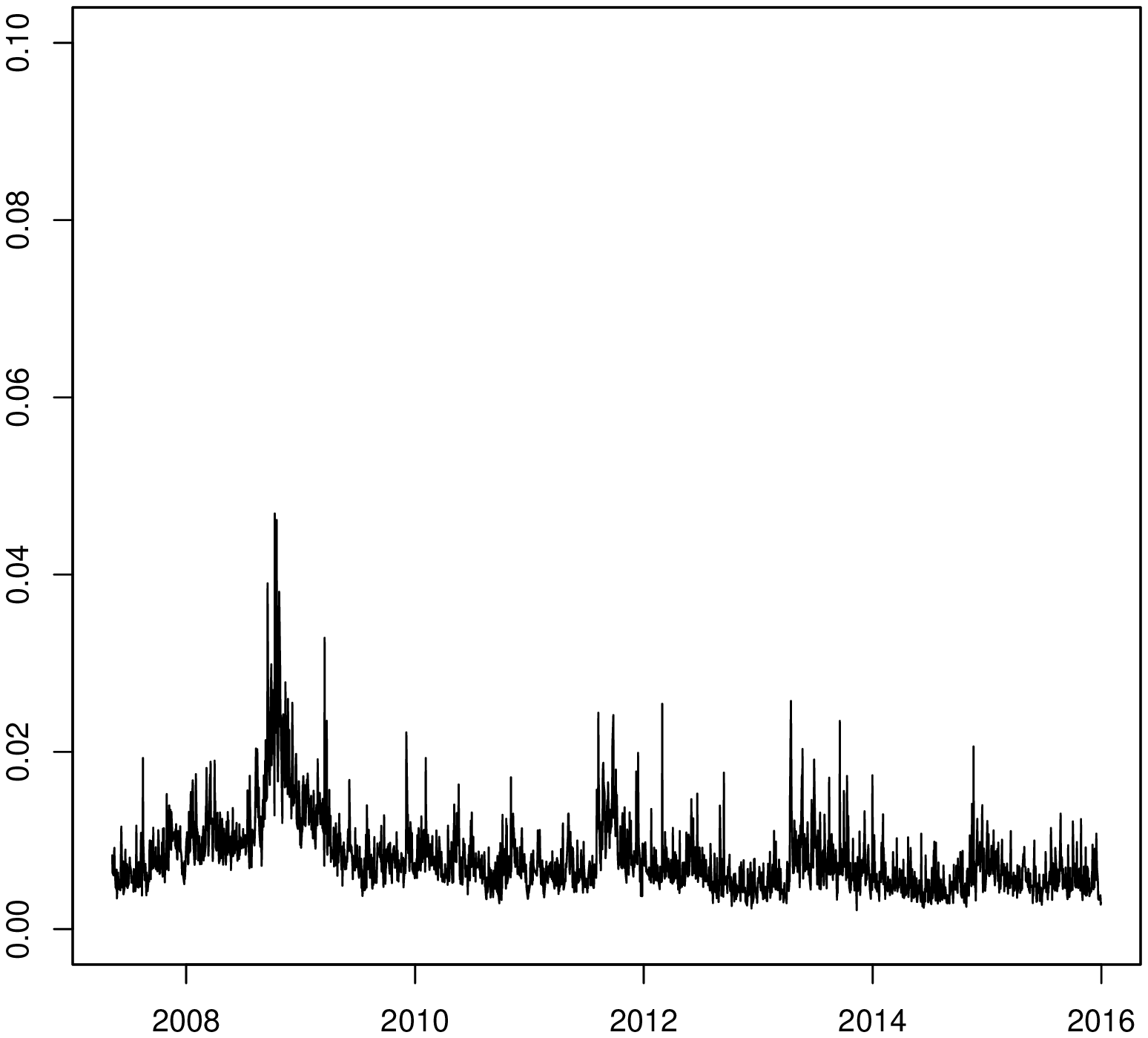}
\end{subfigure}
\begin{subfigure}[b]{0.24\textwidth}
\caption{Copper}\label{fig:RV HG}
\vspace{-10pt}
\includegraphics[width=\textwidth, height=0.8\textwidth]{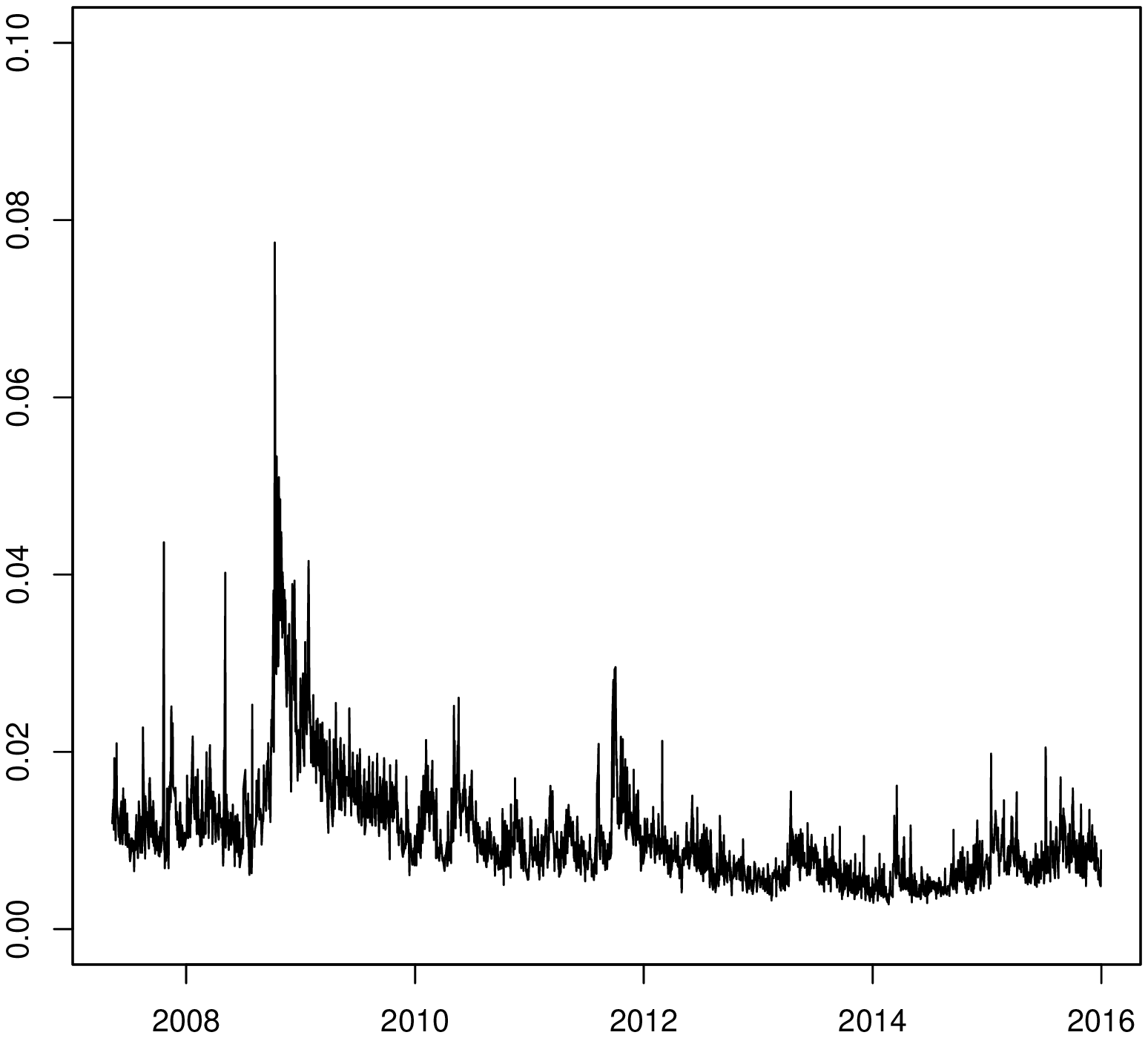}
\end{subfigure}
\begin{subfigure}[b]{0.24\textwidth}
\caption{Natural Gas}\label{fig:RV HG}
\vspace{-10pt}
\includegraphics[width=\textwidth, height=0.8\textwidth]{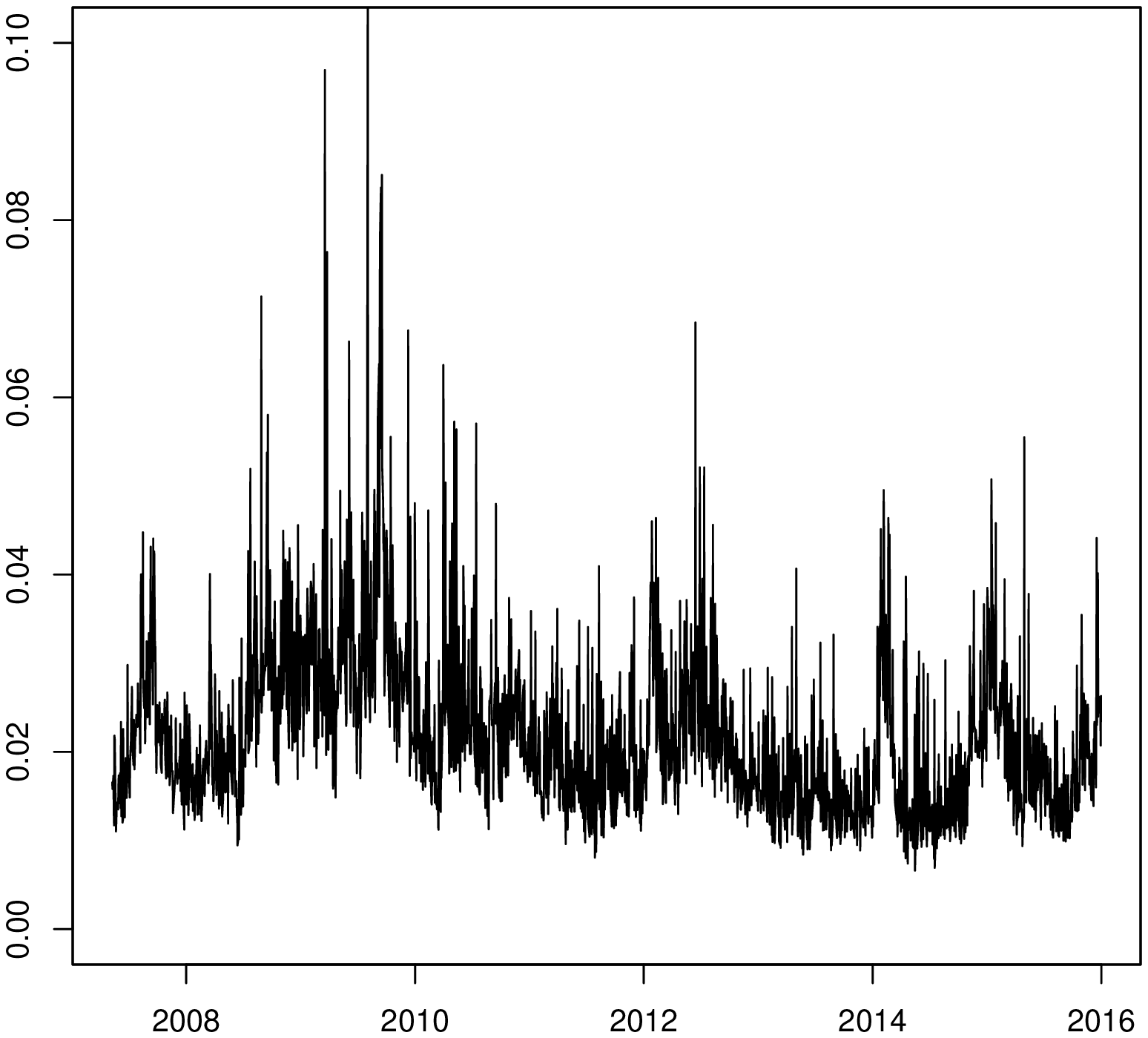}
\end{subfigure}
\begin{subfigure}[b]{0.24\textwidth}
\caption{Silver}\label{fig:RV HG}
\vspace{-10pt}
\includegraphics[width=\textwidth, height=0.8\textwidth]{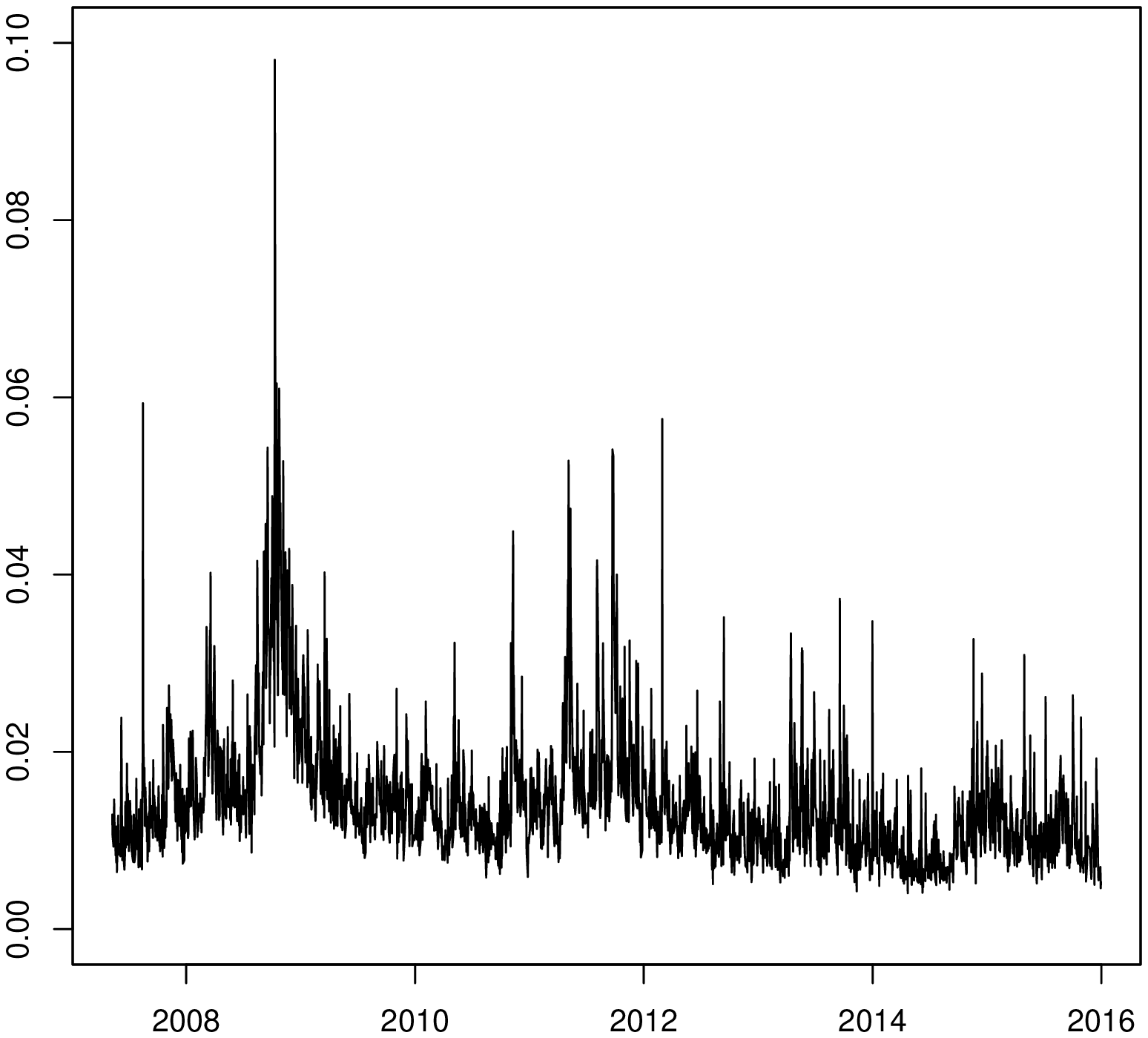}
\end{subfigure}
\captionsetup{justification=centering}
\floatfoot{Note: Plot displays Realized Volatility estimates during the period May 10, 2007 - December 31, 2015.}
\label{fig:Realized Volatility}
\end{figure}

We are interested in the Realized Volatility for the following commodities - Crude Oil (CL), Corn (CN), Cotton (CT), Gold (GC), Copper (HG), Natural Gas (NG), Silver (SV). In the calculation, we consider trades from the period May 10, 2007, until December 31, 2015, during regular trading hours. To ensure sufficient liquidity, we explicitly exclude public holidays as Christmas, Thanksgiving, or Independence Day. From the raw tick data, we extract 5 minutes prices using the last-tick method, and we calculate open-close returns. Additionally, to study the impact of the Global Financial Crisis on the commodity market, we divide our sample into two non--overlapping sub-samples: crisis (10.5.2007 - 9.9.2011) and after--crisis (11.9.2011 - 31.12.2015). Figures \ref{fig: returns}, and \ref{fig:Realized Volatility} show the daily returns and its realized volatility, whereas Table \ref{tab: descriptive stat ret RV} show the descriptive statistics of the data used. 

\begin{table}[ht]
\caption{Descriptive Statistics - Returns, Realized Volatility} 
\centering
\begingroup\small
\begin{tabular}{cccccccc}
  \toprule  & Mean & St.dev. & Skewness & Kurtosis & Median & Minimum & Maximum \\ 
   \cmidrule{1-8} \multicolumn{1}{c}{\textit{Returns}}\\ \cmidrule{1-1}Crude Oil & 0 & 0.0183 & 0.1035 & 4.9358 & 7e-04 & -0.1161 & 0.1434 \\ 
  Corn & 2e-04 & 0.0144 & 0.0607 & 1.9989 & 0 & -0.0653 & 0.086 \\ 
  Cotton & -6e-04 & 0.0145 & -0.2776 & 2.5515 & -3e-04 & -0.0971 & 0.0584 \\ 
  Gold & -1e-04 & 0.009 & 0.0655 & 12.269 & 1e-04 & -0.0728 & 0.1018 \\ 
  Copper & 3e-04 & 0.0118 & -0.3094 & 4.9555 & 5e-04 & -0.0716 & 0.0719 \\ 
  Natural Gas & -9e-04 & 0.0244 & 0.2327 & 2.5368 & -0.0012 & -0.1099 & 0.1507 \\ 
  Silver & -1e-04 & 0.0158 & -0.5099 & 6.034 & 0 & -0.1399 & 0.1031 \\ 
   \cmidrule{1-8} \multicolumn{1}{c}{\textit{Realized Volatility}} &\\ \cmidrule{1-1}Crude Oil & 0.0161 & 0.0087 & 1.9121 & 4.7924 & 0.0139 & 0.0036 & 0.0645 \\ 
  Corn & 0.0132 & 0.0055 & 1.9727 & 8.2438 & 0.012 & 0 & 0.0652 \\ 
  Cotton & 0.0142 & 0.0066 & 1.6882 & 5.6386 & 0.0125 & 0 & 0.0661 \\ 
  Gold & 0.0081 & 0.0043 & 2.5848 & 12.0688 & 0.007 & 0.0021 & 0.0469 \\ 
  Copper & 0.0107 & 0.0062 & 2.6671 & 13.2284 & 0.0093 & 0.0028 & 0.0775 \\ 
  Natural Gas & 0.0217 & 0.0095 & 2.1559 & 9.7943 & 0.0196 & 0.0066 & 0.1117 \\ 
  Silver & 0.0144 & 0.0076 & 2.6149 & 12.8596 & 0.0128 & 0.0041 & 0.0981 \\ 
   \bottomrule\end{tabular}
\endgroup
\captionsetup{justification=centering}
\floatfoot{Note: Table displays descriptive statistics for Returns and Realized Volatility estimates during the period May 10, 2007 - December 31, 2015.}
\label{tab: descriptive stat ret RV}
\end{table}

For the measurement of the ex-ante uncertainty, we use the crude oil (OVX), gold (GVZ) and silver (VXSLV) volatility indexes, commodity counterparts of the CBOE VIX Index. Similar to VIX, commodity indexes measure the  market's expectation of the 30-day volatility using United States Oil Fund, SPDR Gold Shares and Silver ETF options and are calculated according to VIX methodology,\footnote{Full details of the VIX calculation can be found here: \url{http://www.cboe.com/products/vix-index-volatility/vix-options-and-futures/vix-index/the-vix-index-calculation}} i.e. CBOE implied volatility is calculated as

\begin{equation}
\sigma^2_{CBOE}=\frac{2}{T}\sum_{i} \frac{\Delta K_i}{K^2_i}e^{RT}Q(K_i)-\frac{1}{T}\left(\frac{F}{K_0}-1\right)^2,
\end{equation} 
where $T$ is time to expiration; $F$ is the forward index level derived from index option prices; $K_0$ is the first strike below the forward index level $F$; $K_i$ is the strike price of ith out-of-the-money option (call if $K_i > K_0$, put if $K_i< K_0$); $\Delta K_i$ is the interval between strike prices; $R$ is the risk-free interest rate to expiration; and $Q(K_i)$ is the midpoint of the bid-ask spread for each option with strike $K_i$. The value of the Volatility Index that CBOE reports is 
\begin{equation}
Index=100 * \sigma_{CBOE}.
\end{equation} 
Since the CBOE Volatility Indexes report the annual percentage volatility, we construct daily implied volatility by dividing index by $\sqrt{250}$ and 100 to scale it to units of Realized Volatility, e.g. $OVX_{daily}=\frac{\frac{1}{100}*OVX_{annual}}{\sqrt{250}}$.

\begin{figure}[H]
\centering
\caption{Daily CBOE Volatility Indexes} 
\begin{subfigure}[b]{0.24\textwidth}
\caption{Crude Oil}\label{fig:INDEX CL}
\vspace{-10pt}
\includegraphics[width=\textwidth, height=0.8\textwidth]{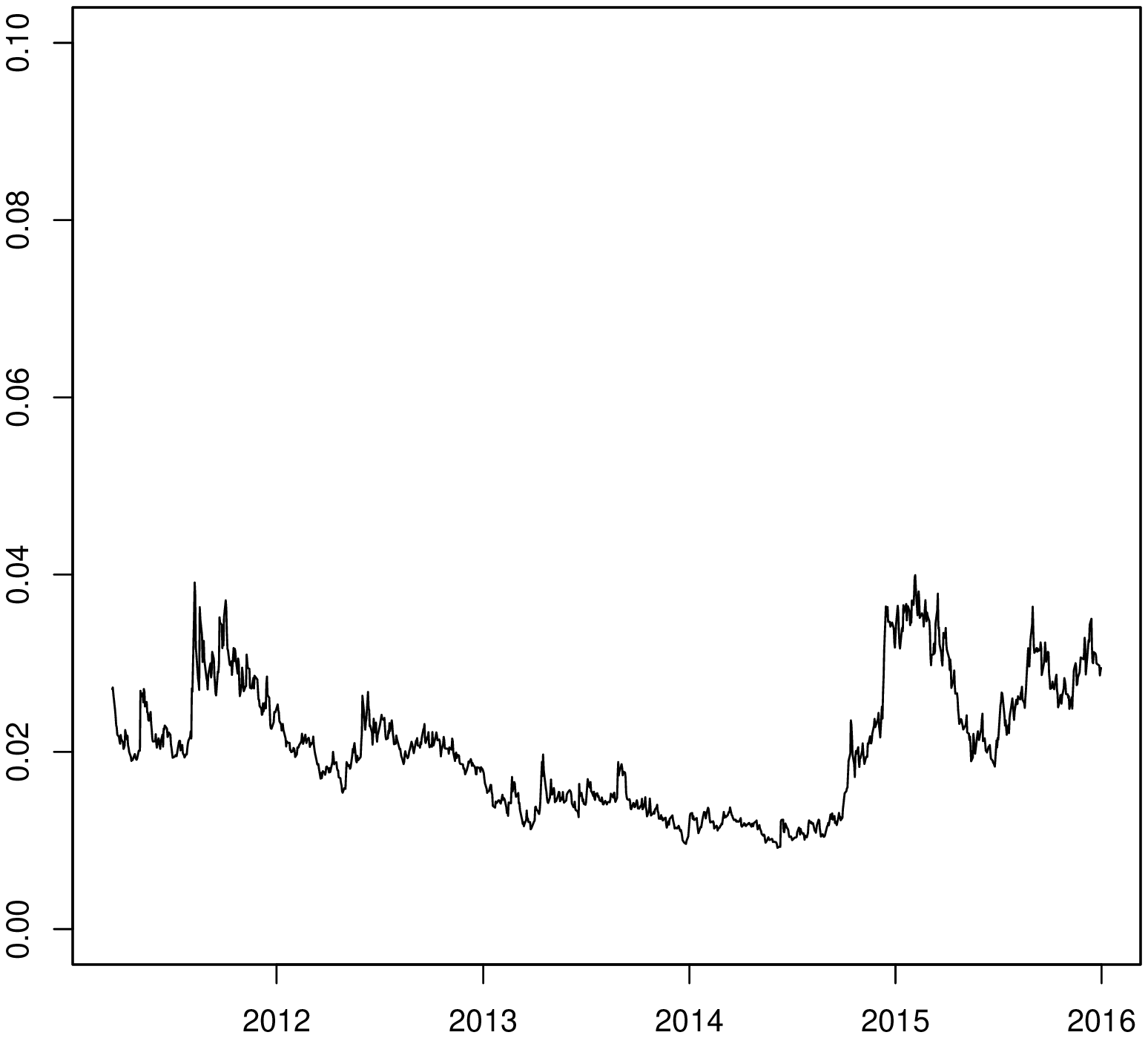}
\end{subfigure}
\begin{subfigure}[b]{0.24\textwidth}
\caption{Gold}\label{fig:INDEX GC}
\vspace{-10pt}
\includegraphics[width=\textwidth, height=0.8\textwidth]{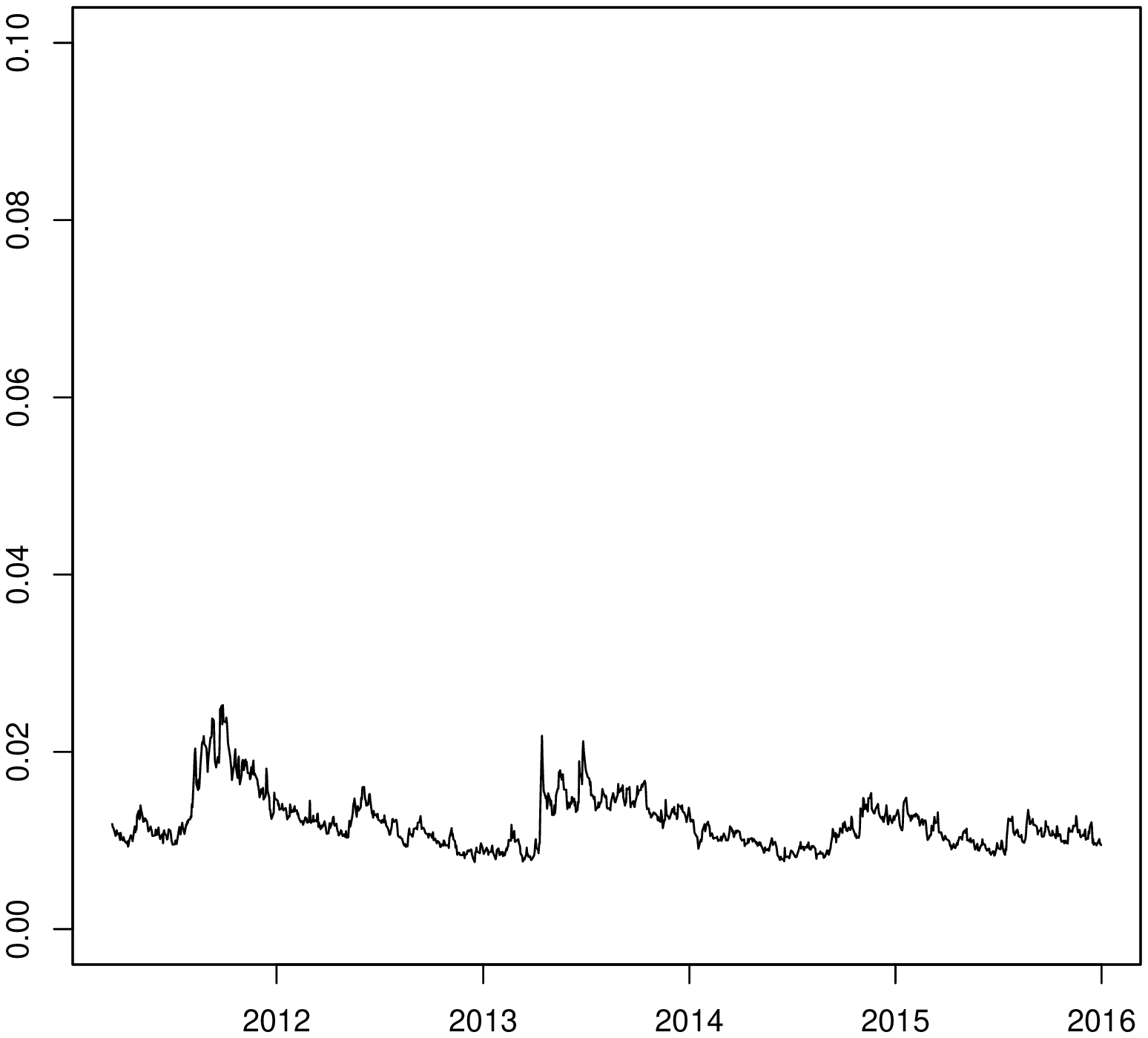}
\end{subfigure}
\begin{subfigure}[b]{0.24\textwidth}
\caption{Silver}\label{fig:INDEX HG}
\vspace{-10pt}
\includegraphics[width=\textwidth, height=0.8\textwidth]{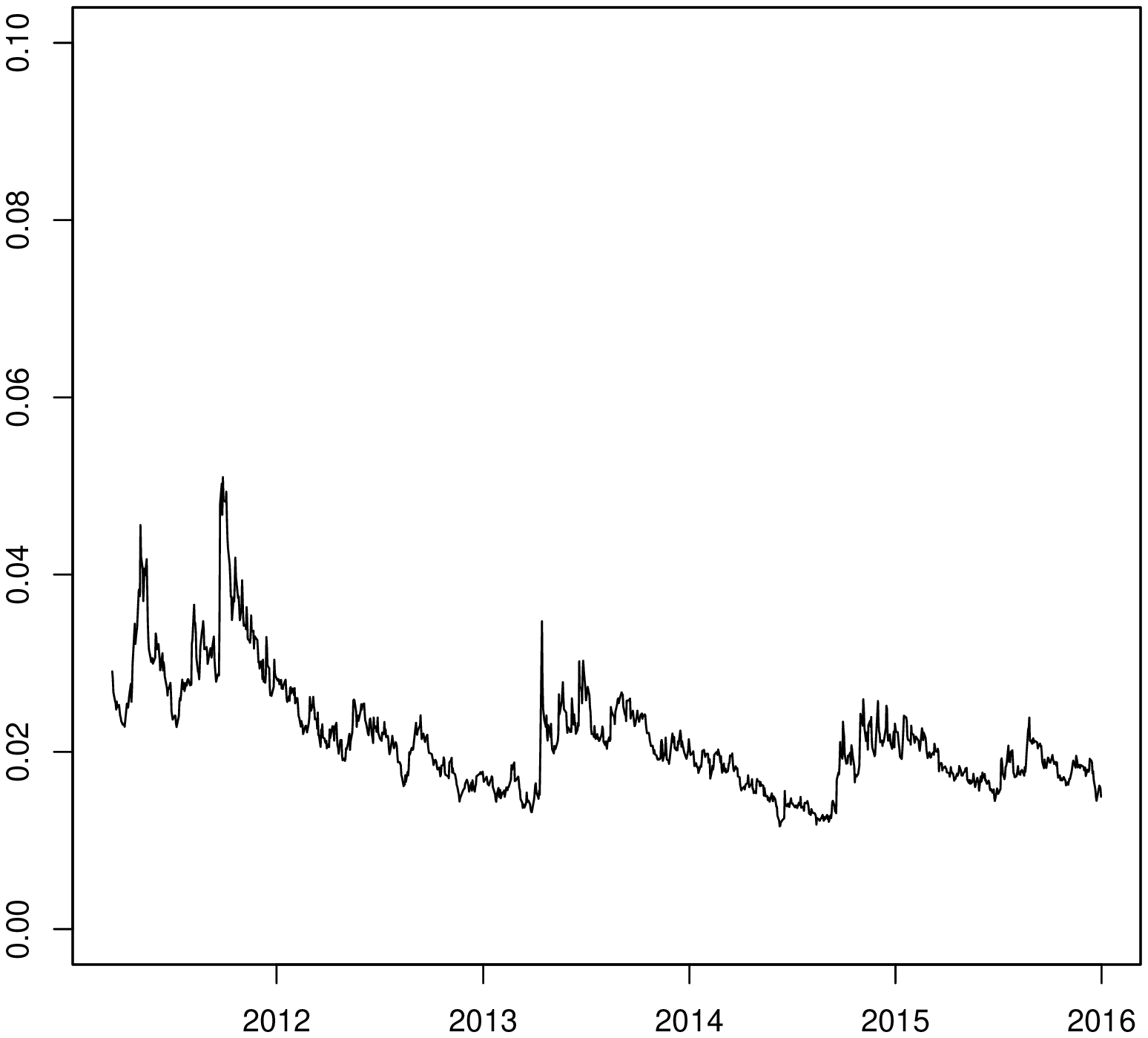}
\end{subfigure}
\captionsetup{justification=centering}
\floatfoot{Note: Plot displays daily CBOE Volatility Indexes during period March 16, 2011 - December 31, 2015.}
\label{fig:CBOE Volatility Indexes}
\end{figure}

The Crude oil, Gold, and Silver CBOE Volatility indices are obtained from the Federal Reserve Bank at Saint Louis.\footnote{\url{https://fred.stlouisfed.org/categories/32425}} Data availability differs across commodities with Crude Oil having the longest history\footnote{The OVX was officially launched on July 15, 2008, but values were calculated back to May 10, 2007, when CBOE began trading the United States Oil Fund options.}, since May 10, 2007; Gold being available from June 3, 2008; and Silver from March 16, 2011. To have a balanced panel in our application, we set the starting date of all indexes to March 16, 2011. Figure \ref{fig:CBOE Volatility Indexes} shows the plot of the volatiltiy indexes, and Table \ref{tab: descriptive stat INDEX} descriptive statistics.

\begin{table}[ht]
\caption{Descriptive Statistics - CBOE Volatility Indexes}
\centering
\begingroup\small
\begin{tabular}{Cccccccc}
  \toprule  & Mean & St.dev. & Skewness & Kurtosis & Median & Minimum & Maximum \\ 
   \cmidrule{2-8}Crude Oil - OVX & 0.0204 & 0.0071 & 0.5016 & -0.6084 & 0.0201 & 0.0092 & 0.0399 \\ 
  Gold - GVZ & 0.012 & 0.003 & 1.3722 & 2.3039 & 0.0113 & 0.0076 & 0.0253 \\ 
  Silver - VXSLV & 0.0215 & 0.0062 & 1.4368 & 2.9751 & 0.0204 & 0.0116 & 0.051 \\ 
   \bottomrule\end{tabular}
\endgroup
\captionsetup{justification=centering}
\floatfoot{Note: Table displays descriptive statistics of CBOE Volatility Indexes during period March 16, 2011 - December 31, 2015.}
\label{tab: descriptive stat INDEX}
\end{table}

\section{Empirical Application}\label{empir app}
In the first part of the empirical analysis, we study information content of the Realized Volatility for the commodity Value--at--Risk estimation. Specifically, we concentrate on the commonalities in the dependence of the VaRs and corresponding Realized Volatilities during the period of the Global Financial Crisis. The second part complements the Realized Volatility analysis and study also role of the commodity CBOE Volatility Indexes for the Value--at--Risk estimation.  

\subsection{Influence of Realized Volatility}\label{sec: Realized Volatility}

To study the Value-at-Risk of commodities, we propose the following panel quantile regression model that links future return quantiles with past Realized Volatility as
\begin{equation}\label{eq: VaR RV}
Q_{r_{i,t+1}}(\tau)= \alpha_{i}(\tau) + \beta_{RV^{1/2}}(\tau)*RV_{i,t}^{1/2},
\end{equation}
where $i\in \left\{CL, CN, CT, GC, HG, NG, SV \right\}$. The appealing features of the model described by the \autoref{eq: VaR RV} are a possibility to identify common patterns in Realized Volatilities by controlling for unobserved heterogeneity among commodities; and directly relate Value--at--Risk and Realized Volatility as represented by basic definition of parametric VaR in \cite{Longerstaey1996}. It is important to highlight the fact that we do not need to assume the parametric distribution. 

\begin{figure}[H]
\centering
\caption{Parameter estimates: Realized Volatility} 
\begin{subfigure}[b]{0.325\textwidth}
\caption{\textit{crisis}}\label{fig:PQR-RV crisis}
\vspace{-20pt}
\includegraphics[width=\textwidth, height=1.2\textwidth]{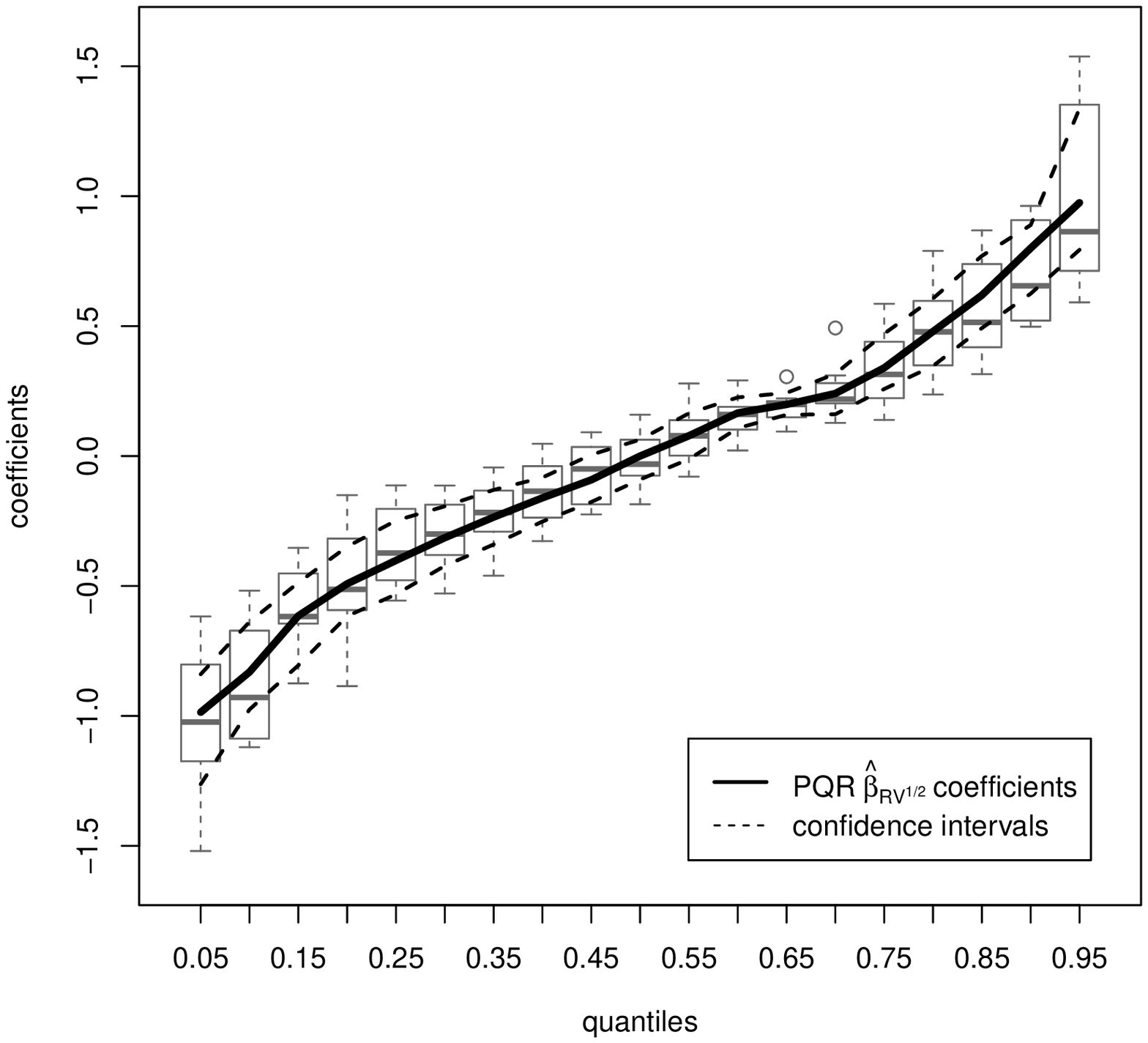}
\end{subfigure}
\begin{subfigure}[b]{0.325\textwidth}
\caption{\textit{after crisis}}\label{fig:PQR-RV after}
\vspace{-20pt}
\includegraphics[width=\textwidth, height=1.2\textwidth]{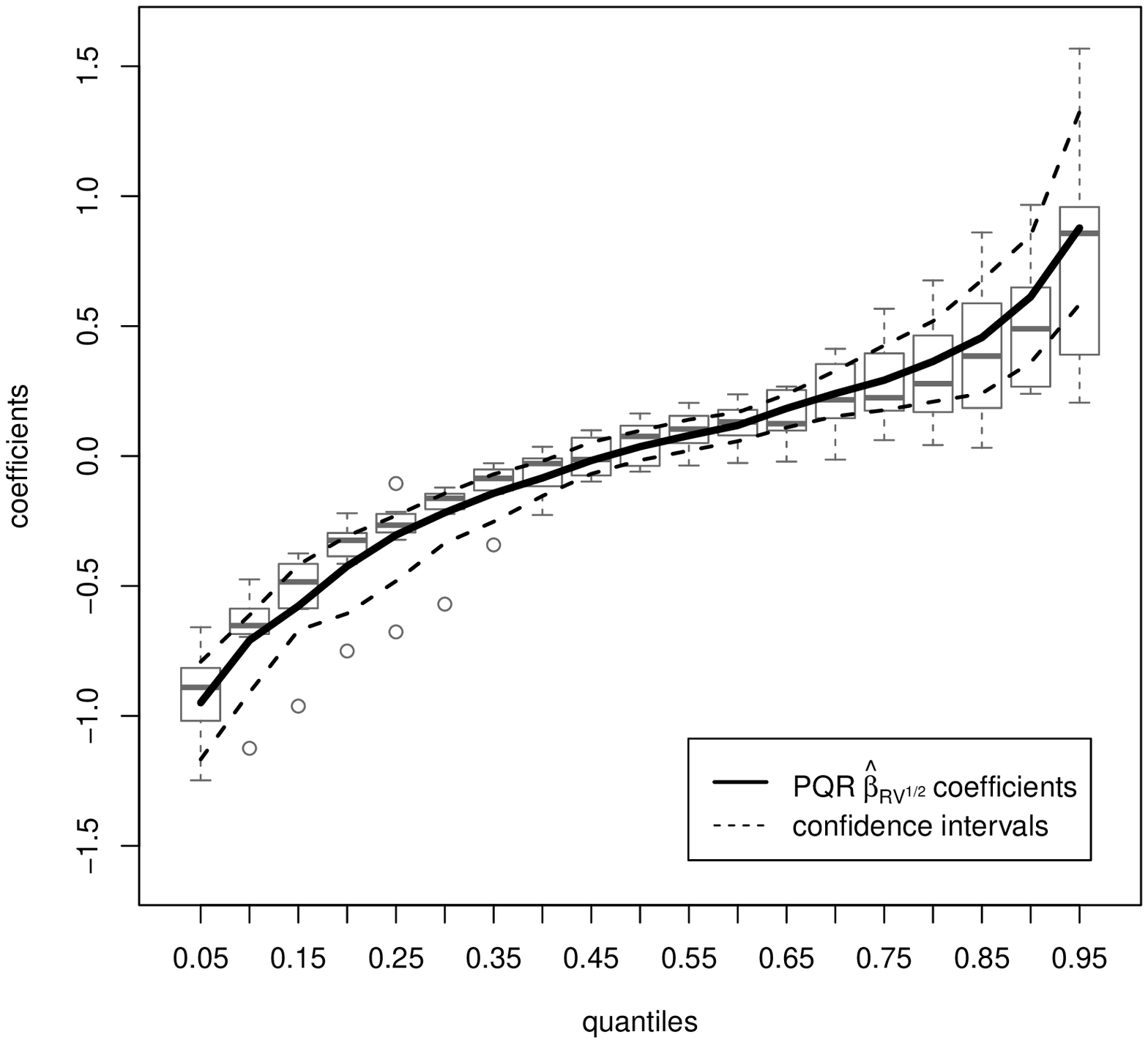}
\end{subfigure}
\begin{subfigure}[b]{0.325\textwidth}
\caption{\textit{full sample}}\label{fig:PQR-RV full}
\vspace{-20pt}
\includegraphics[width=\textwidth, height=1.2\textwidth]{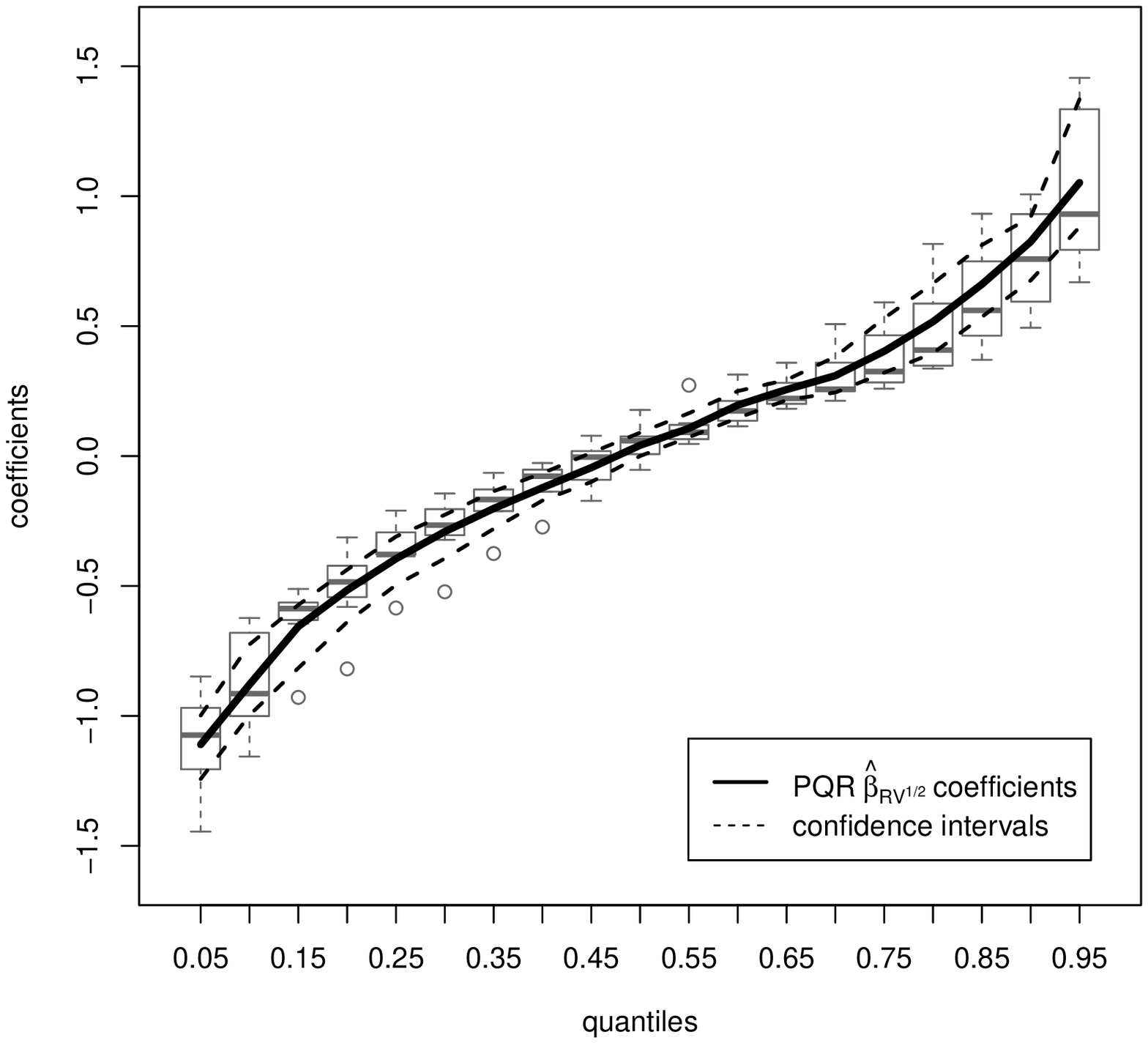}
\end{subfigure}
\vspace{-10pt}
\captionsetup{justification=centering}
\floatfoot{Note: Parameters estimates with corresponding 95\% confidence intervals are plotted by solid and dashed lines respectively. Individual univariate estimates are plotted in boxplots.}
\label{fig:PQR-RV all commodity}
\end{figure}

\begin{figure}[H]
\centering
\caption{Conditional returns distributions vs standard normal distribution} 
\vspace{-20pt}
\includegraphics[scale=0.4]{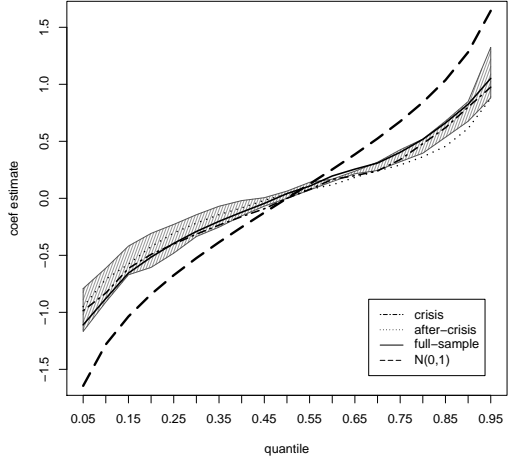}
\vspace{-10pt}
\captionsetup{justification=centering}
\floatfoot{Note: Hatched area represents intersection of 95\% confidence intervals of all studied samples estimates.}
\label{fig:PQR-RV vs normal}
\end{figure}

\autoref{tab:PQR RV all commodity}, \autoref{fig:PQR-RV all commodity} and \autoref{fig:PQR-RV vs normal} summarize the first part of our results. We identify strong common patterns of the dependence between quantiles of future commodity returns and ex-post Realized Volatility and find this dependence to be highly statistically significant across the whole return distribution. Specifically, the conditional returns distribution share commonalities within the group of selected commodities and these commonalities are stable over time. 

For the clarity and better readability results presented in \autoref{tab:PQR RV all commodity} are divided into three groups according to period included in the analysis. In the first part, we analyze the role of the Global Financial Crisis in the commodity markets; in the second part after-crisis period; and the last part provides estimates covering the full dataset. 

\begin{table}[H]
\begin{center}
\small
\caption{Parameter estimates: Realized Volatility} \label{tab:PQR RV all commodity}
\begin{tabular}{lccccccc}
\toprule
$\tau$ & 5\% & 10\% & 25\% & 50\% & 75\% & 90\% & 95\% \\ \cmidrule{2-8} \\[-0.75em]
 & \multicolumn{7}{c}{\textit{crisis: 10.5.2007 - 9.9.2011}} \\ \cmidrule{2-8}
$\hat{\beta}_{RV^{1/2}}$ & -0.986 & -0.831 & -0.402 & 0 & 0.339 & 0.8 & 0.975 \\ 
 & (-9.13) & (-8.39) & (-6.07) & (0) & (6.13) & (10.18) & (6.91) \\  \cmidrule{2-8} \\[-0.75em]
 & \multicolumn{7}{c}{\textit{after crisis: 11.9.2011 - 31.12.2015}} \\ \cmidrule{2-8} 
$\hat{\beta}_{RV^{1/2}}$ & -0.949 & -0.709 & -0.304 & 0.036 & 0.292 & 0.613 & 0.878 \\ 
 & (-11.05) & (-9.17) & (-4.38) & (1.23) & (4.99) & (5.12) & (4.91) \\  \cmidrule{2-8} \\[-0.75em]
 & \multicolumn{7}{c}{\textit{full sample: 10.5.2007 - 31.12.2015}} \\ \cmidrule{2-8} 
$\hat{\beta}_{RV^{1/2}}$ & -1.11 & -0.879 & -0.395 & 0.041 & 0.403 & 0.825 & 1.052 \\ 
 & (-15.61) & (-11.43) & (-8.32) & (1.76) & (7.1) & (12.26) & (8.11) \\  
\bottomrule
\end{tabular}
\begin{tablenotes}
\centering
\footnotesize
\item{Note: Table displays coefficient estimates with bootstraped t-statistics in parentheses. Full results with the individual fixed effects $\alpha_i(\tau)$ are presented in \nameref{appendix}}
\end{tablenotes}
\end{center}
\end{table}

Throughout the whole \autoref{tab:PQR RV all commodity}, we can see the high statistical significance of the  $\hat{\beta}_{RV^{1/2}}$ coefficient estimates for all but median quantiles. Median performance, especially lack of the explanatory power to model median returns, constitutes stylized fact of the unpredictability of expected returns and is in line with Efficient Market Hypothesis \citep{fama1970efficient}.    

Turning our attention to remaining quantiles, we can observe minor differences in the relative influence of the Realized Volatility on the Value--at--Risk estimation within studied periods. During the turbulent times of financial crisis, the role of the Realized Volatility is slightly more important compared to the after-crisis period. We draw this conclusion since the absolute values of coefficient estimates are always higher during crisis time, e.g. during crisis 95\% quantile coefficient estimate of 0.975 vs after crisis value of 0.878. Our results confirm \textit{Stylized Fact 6} \citep{christoffersen2018factor} about increased uncertainty in the commodity markets during times of financial distress. Moreover, the absolute values of the $\hat{\beta}_{RV^{1/2}}$ estimates are higher in lower than in upper quantiles.  We document the relatively higher influence of the ex-post uncertainty on the downside risk estimation. This asymmetry in the lower/upper quantiles of the conditional returns is similar to unconditional gain/loss asymmetry, \textit{Stylized Fact 3} defined in \cite{cont2001empirical}.

An interesting finding of our analysis is the shape of the conditional returns distribution that can be seen from the graphical representation of the parameter estimates in the \autoref{fig:PQR-RV all commodity} and \autoref{fig:PQR-RV vs normal}. In the \autoref{fig:PQR-RV all commodity}, we can see similar shapes of the common conditional returns distribution represented by panel quantile regression coefficient estimates (solid black lines) for all studied commodities. These similarities are visible also in the level of individual commodities represented by boxplots. A closer look on the tails of the distributions reveals slight asymmetric influence in the lower and upper tails. 
Importantly, \autoref{fig:PQR-RV vs normal} contrasts the parameter estimates with the Standard Normal Distribution and shows platykurtic conditional return distribution since both lower and upper tails are thinner than those implied by Standard Normal Distribution. Moreover, in the lower tail, conditional return distributions of all samples lies in the intersection of corresponding 95\% confidence intervals. Thus we conclude, that the properties and characteristics of the commodity market downside risk are stable over time and did not change significantly during the Global Financial Crisis.

These findings are in contrast to parametric studies, where fat-tailed distributions were required to match empirical data, we document that returns standardized by the Realized Volatility have thin tails and are \textit{platykurtic}. Although our results are in sharp contrast to the general perception of returns behavior, they are in line with works that documents diversification benefits of the commodities \citep{belousova2012diversification}, and consider commodities to be less risky than stocks \citep{bodie1980risk, gorton2006facts, conover2010now}. Moreover, if we combine lower riskiness of commodities with results of \cite{andersen2000exchange} where conditional returns of financial assets are documented to be almost Gaussian, conditional commodity returns should indeed have thinner tails than financial assets.

\subsection{Role of ex-ante uncertainty} 

Revealing the importance of ex-post uncertainty for future commodity Value--at--Risk, and identifying common patterns in the panel of commodities, it is tempting to ask what is the role of ex-ante information. In this section, we extend our analysis and study also role of the ex-ante implied volatility measure. Since the availability of the commodity CBOE Volatility Indexes is limited, period used for study spans from March 16, 2011, to December 31, 2015, and almost corresponds to after--crisis period from the previous section. 

Similar to the previous section, we concentrate on the role of the ex-post or ex-ante volatility measure for the Value--at--Risk estimation in line with the parametric definition of VaR. In addition to the Realized Volatility, we also use implied volatility index to explain the future conditional returns, hence we estimate following two equations and contrast the parameters first 
\begin{equation}\label{eq: VaR RV - RV + INDEX}
Q_{r_{i,t+1}}(\tau)= \alpha_{i}(\tau) + \beta_{RV^{1/2}}(\tau)*RV_{i,t}^{1/2},
\end{equation}
\begin{equation}\label{eq: VaR INDEX - RV + INDEX}
Q_{r_{i,t+1}}(\tau)= \alpha_{i}(\tau) + \beta_{INDEX^{1/2}}(\tau)*INDEX_{i,t}^{1/2},
\end{equation}
While both approaches result in semi-parametric VaR and give us conditional returns distribution, the later one stresses the importance of the anticipated risk level of the market participants.  

In the last part of our analysis, we examine information content of ex-ante uncertainty after controlling for ex-post uncertainty, and we formulate the problem as   
\begin{equation}\label{eq: VaR RV + INDEX - RV + INDEX}
Q_{r_{i,t+1}}(\tau)= \alpha_{i}(\tau) + \beta_{RV^{1/2}}(\tau)*RV_{i,t}^{1/2} + \beta_{INDEX^{1/2}}(\tau)*INDEX_{i,t}^{1/2}.
\end{equation}
This specification allows us to directly compare role and importance of the ex-post and ex-ante risk measures in the determination of quantiles of future returns. Equations \autoref{eq: VaR RV - RV + INDEX}, \autoref{eq: VaR INDEX - RV + INDEX} and \autoref{eq: VaR RV + INDEX - RV + INDEX} are estimated on the panel of $i\in \left\{CL, GC, SV \right\}$ commodities.

\begin{table}[H]
\begin{center}
\caption{Parameter estimates: Realized Volatility \& CBOE Volatility Index} \label{tab:PQR RV INDEX commodity}
\begin{tabular}{LCCCCCCC}
\toprule
\rowfont{\small}
$\tau$ & 5\% & 10\% & 25\% & 50\% & 75\% & 90\% & 95\% \\ \cmidrule{1-8} \\[-0.75em]
\textit{Panel A} &  \\
& \multicolumn{7}{c}{$Q_{r_{i,t+1}}(\tau)= \alpha_{i}(\tau) + \beta_{RV^{1/2}}(\tau)*RV_{i,t}^{1/2}$} \\ \cmidrule{2-8}
$\hat{\beta}_{RV^{1/2}}$ & -1.167 & -0.884 & -0.392 & 0.098 & 0.438 & 0.786 & 0.994 \\ 
&  (-65.86) & (-8.09) & (-5.76) & (4.46) & (5.86) & (5.24) & (3.41) \\ 
\rowfont{\footnotesize}
$\hat{\alpha}_{CL}$ & -0.009 & -0.005 & -0.003 & 0 & 0.002 & 0.006 & 0.007 \\ 
&  (-28.76) & (-3.47) & (-3.68) & (-2.03) & (2.33) & (3.19) & (2.12) \\
$\hat{\alpha}_{GC}$ & -0.004 & -0.002 & -0.001 & -0.001 & 0.001 & 0.003 & 0.005 \\ 
&  (-30.49) & (-3.5) & (-3.09) & (-5.28) & (1.22) & (3.45) & (2.71) \\ 
$\hat{\alpha}_{SV}$ & -0.007 & -0.004 & -0.002 & -0.001 & 0.001 & 0.005 & 0.009 \\ 
&  (-39.52) & (-3.62) & (-2.76) & (-6.42) & (1.31) & (2.88) & (2.99) \\ \cmidrule{2-8} \\[-0.75em]
\rowfont{\small}
& \multicolumn{7}{c}{$Q_{r_{i,t+1}}(\tau)= \alpha_{i}(\tau) + \beta_{INDEX^{1/2}}(\tau)*INDEX_{i,t}^{1/2}$} \\ \cmidrule{2-8} 
$\hat{\beta}_{INDEX^{1/2}}$ & -1.229 & -0.859 & -0.404 & 0.098 & 0.449 & 0.821 & 1.153 \\ 
&  (-18.07) & (-15.37) & (-21.47) & (3.78) & (30.75) & (14.19) & (17.06) \\ 
\rowfont{\footnotesize}
$\hat{\alpha}_{CL}$ & 0.001 & 0 & 0 & -0.001 & -0.001 & -0.001 & -0.002 \\ 
&  (0.41) & (0.5) & (-0.52) & (-2.02) & (-5.32) & (-1.09) & (-1.77) \\
$\hat{\alpha}_{GC}$ & 0.003 & 0.002 & 0.001 & -0.001 & -0.002 & -0.001 & -0.003 \\ 
&  (4.2) & (2.8) & (4.38) & (-4.46) & (-12.03) & (-2.44) & (-4.17) \\
$\hat{\alpha}_{SV}$ & 0.004 & 0.003 & 0.002 & -0.002 & -0.003 & -0.003 & -0.003 \\ 
&  (3.59) & (2.84) & (3.91) & (-4.72) & (-9.57) & (-2.44) & (-2.42) \\ \cmidrule{1-8} \\[-0.75em]
\rowfont{\small}
\textit{Panel B} &  \\
 & \multicolumn{7}{c}{$Q_{r_{i,t+1}}(\tau)= \alpha_{i}(\tau) + \beta_{RV^{1/2}}(\tau)*RV_{i,t}^{1/2} + \beta_{INDEX^{1/2}}(\tau)*INDEX_{i,t}^{1/2}$} \\ \cmidrule{2-8}
$\hat{\beta}_{RV^{1/2}}$ & -0.465 & -0.299 & -0.159 & 0.033 & 0.148 & 0.19 & 0.02 \\ 
 & (-3.69) & (-7.35) & (-1.85) & (0.95) & (3.07) & (1.56) & (0.1) \\  
$\hat{\beta}_{INDEX^{1/2}}$ & -0.838 & -0.654 & -0.281 & 0.083 & 0.355 & 0.678 & 1.131 \\ 
 & (-5.14) & (-8.16) & (-6.2) & (3.22) & (18.08) & (16.59) & (7.99) \\ 
\rowfont{\footnotesize}
$\hat{\alpha}_{CL}$ & -0.001 & 0.001 & -0.001 & -0.001 & -0.001 & -0.001 & -0.002 \\ 
 & (-0.54) & (0.71) & (-1.55) & (-2.5) & (-7.17) & (-0.9) & (-1.67) \\
$\hat{\alpha}_{GC}$ & 0.001 & 0.001 & 0.001 & -0.001 & -0.002 & -0.001 & -0.003 \\ 
 & (1.07) & (1.77) & (3.76) & (-5.17) & (-17.94) & (-2.58) & (-3.75) \\ 
$\hat{\alpha}_{SV}$ & 0.002 & 0.002 & 0.001 & -0.002 & -0.003 & -0.002 & -0.003 \\ 
 & (1.36) & (2.18) & (3.27) & (-4.82) & (-11.01) & (-2.75) & (-2.36) \\ 
\bottomrule
\end{tabular}
\begin{tablenotes}
\centering
\footnotesize
\item{Note: Table displays coefficient estimates with bootstraped t-statistics in parentheses.}
\end{tablenotes}
\end{center}
\end{table}

\begin{figure}[H]
\centering
\caption{Conditional returns distributions vs Standard normal distribution} 
\vspace{-20pt}
\includegraphics[scale=0.4]{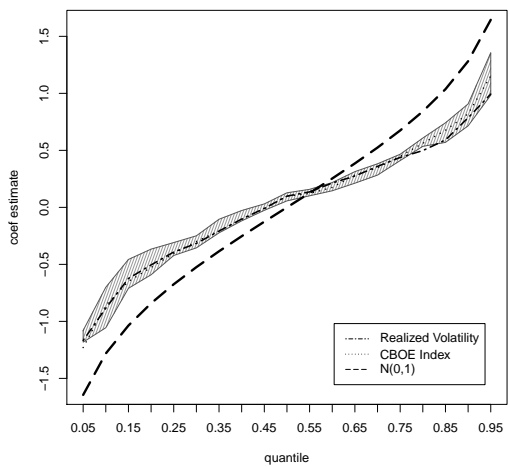}
\vspace{-10pt}
\captionsetup{justification=centering}
\floatfoot{Note: Hatched area represents intersection of 95\% confidence intervals of separate estimation of ex-post and ex-ante volatility based models.}
\label{fig:PQR-RV INDEX vs normal}
\end{figure}

The \textit{Panel A} of the \autoref{tab:PQR RV INDEX commodity} highlight importance of both individual Realized Volatility and CBOE Volatility Indexes for future commodity returns quantiles modelling. According to the proximity of the conditional return distributions represented by the coefficient estimates $\hat{\beta}_{RV^{1/2}}$ and $\hat{\beta}_{INDEX^{1/2}}$, the impact of both ex-post and ex-ante volatility on future return quantiles is of similar extent. We support this finding by visual inspection of \autoref{fig:PQR-RV INDEX vs normal} - the conditional return distributions from the ex-post and ex-ante volatility model specification are close to each other, importantly, they both lie in the intersection of the 95\% confidence intervals of these estimates. 

In the \textit{Panel A}, we further see the high statistical significance of all $\hat{\beta}_{RV^{1/2}}$ and $\hat{\beta}_{INDEX^{1/2}}$ coefficients, including the median. In contrast to our previous analysis, the median individual fixed effect estimates $\alpha_i$ are also statistically significant. Since the signs of $\alpha$ and $\beta$ are opposite, they offset the influence of each other and make median returns difficult to predict. Therefore return distributions conditional on either of the volatility measures qualitatively match the results of our previous analysis. There is an asymmetric influence of the volatility measures in the lower and upper quantiles. Greater asymmetry is present in $\hat{\beta}_{RV^{1/2}}$ coefficients where the difference between 5\% and 95\% quantile values is 1.167-0.994= 0.173 compared to $\hat{\beta}_{INDEX^{1/2}}$ difference of 1.229-1.153=0.076. We also document that conditional return distributions have thinner tails than the Standard Normal Distribution and therefore are \textit{platykurtic}. This characteristic is visible in the \autoref{fig:PQR-RV INDEX vs normal} where the 5\% (95\%) conditional return distribution quantile of Realized Volatility specification is -1.167 (0.994) opposed to -1.645 (1.645) of Standard Normal Distribution. Similarly for CBOE Index where 5\% (95\%) conditional return distribution quantile is -1.229 (1.153). 

\begin{figure}[H]
\centering
\caption{Parameter estimates: Realized Volatility + CBOE VOlatility Index} 
\begin{subfigure}[b]{0.85\textwidth}
\caption{\textit{RV + CBOE Index}}\label{fig:PQR RV Index RV commodity}
\vspace{-20pt}
\includegraphics[scale=0.4]{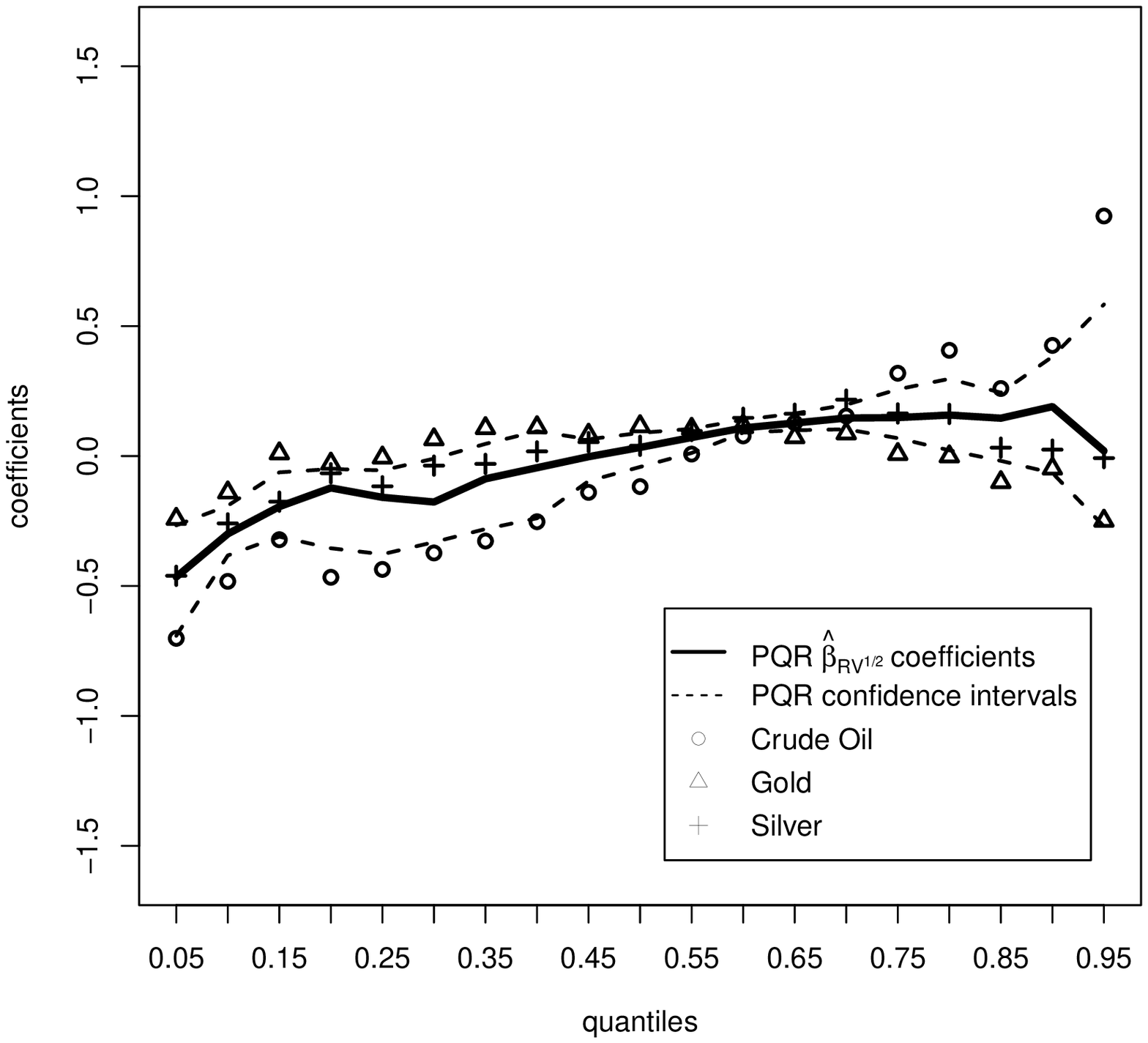}
\includegraphics[scale=0.4]{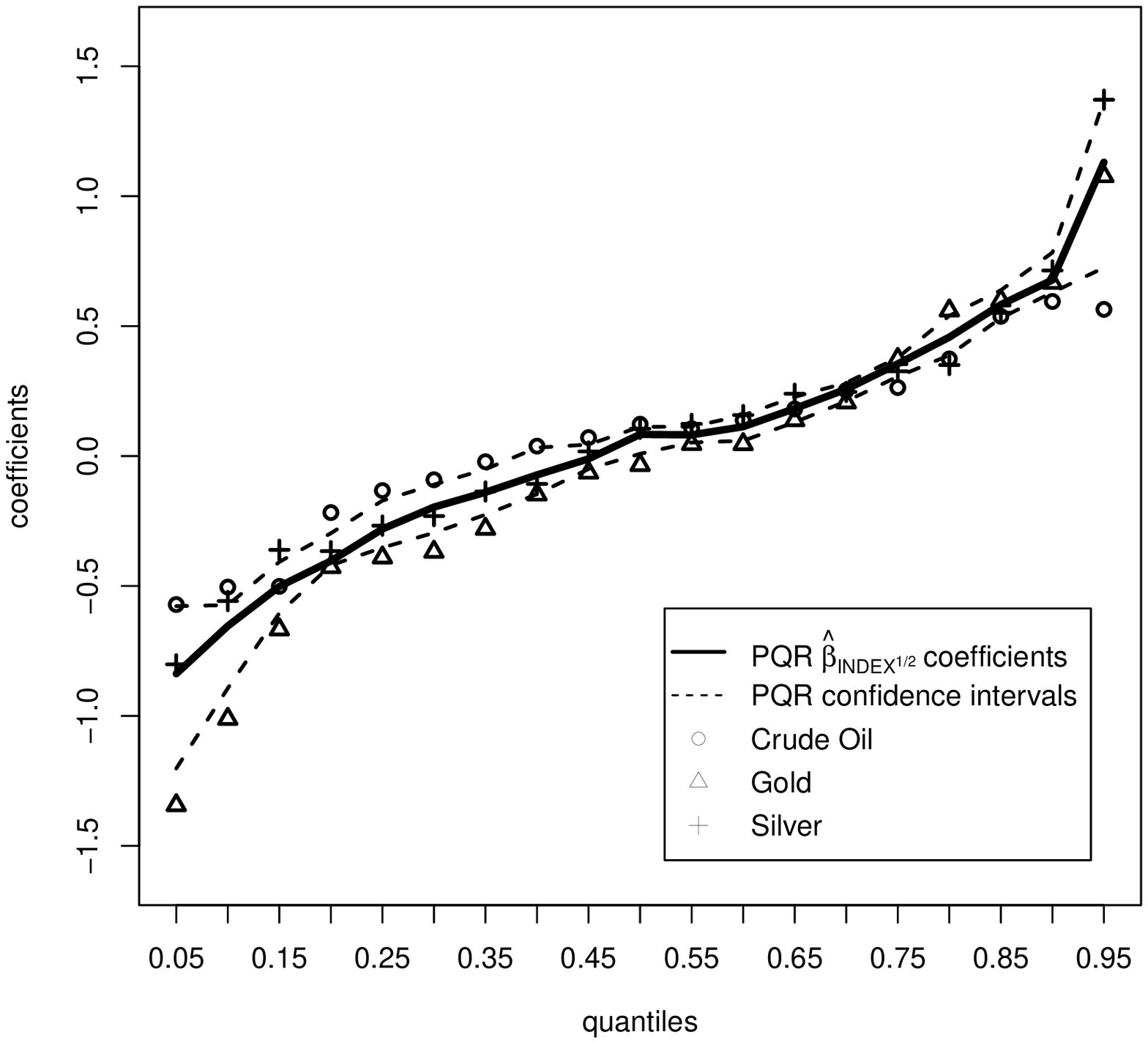}
\end{subfigure}
\captionsetup{justification=centering}
\floatfoot{Note: For both realized volatility and volatility index parameters estimates with corresponding 95\% confidence intervals are ploted by solid and dashed lines respectively. Individual univariate estimates are indicated by circle (Crude Oil), triangle (Gold), plus sign (Silver).}
\label{fig:PQR RV INDEX}
\end{figure}

\textit{Panel B} of the \autoref{tab:PQR RV INDEX commodity} emphasizes the role of ex-ante volatility in the Value--at--Risk estimation. The CBOE Volatility Indexes show great importance once we control for Realized Volatility. The dominance of the CBOE Volatility Indexes is well documented in the above median quantiles - the $\hat{\beta}_{INDEX^{1/2}}$ estimates are always statistically significant while it is not generally true for $\hat{\beta}^{RV^{1/2}}$ coefficients. In far upper quantiles, $\hat{\beta}_{RV^{1/2}}$ coefficient estimates are even not statistically different from zero. \autoref{fig:PQR RV INDEX} confirms our findings visually. The confidence band of the Realized Volatility coefficients are much wider than confidence intervals of the CBOE Volatility Indexes coefficients and often contains 0. 

\begin{figure}[H]
\centering
\caption{Parameter estimates: individual Realized Volatility \& CBOE Volatility Index} 
\begin{subfigure}[b]{0.425\textwidth}
\caption{\textit{RV}}\label{fig:PQR RV only commodity}
\vspace{-20pt}
\includegraphics[scale=0.4]{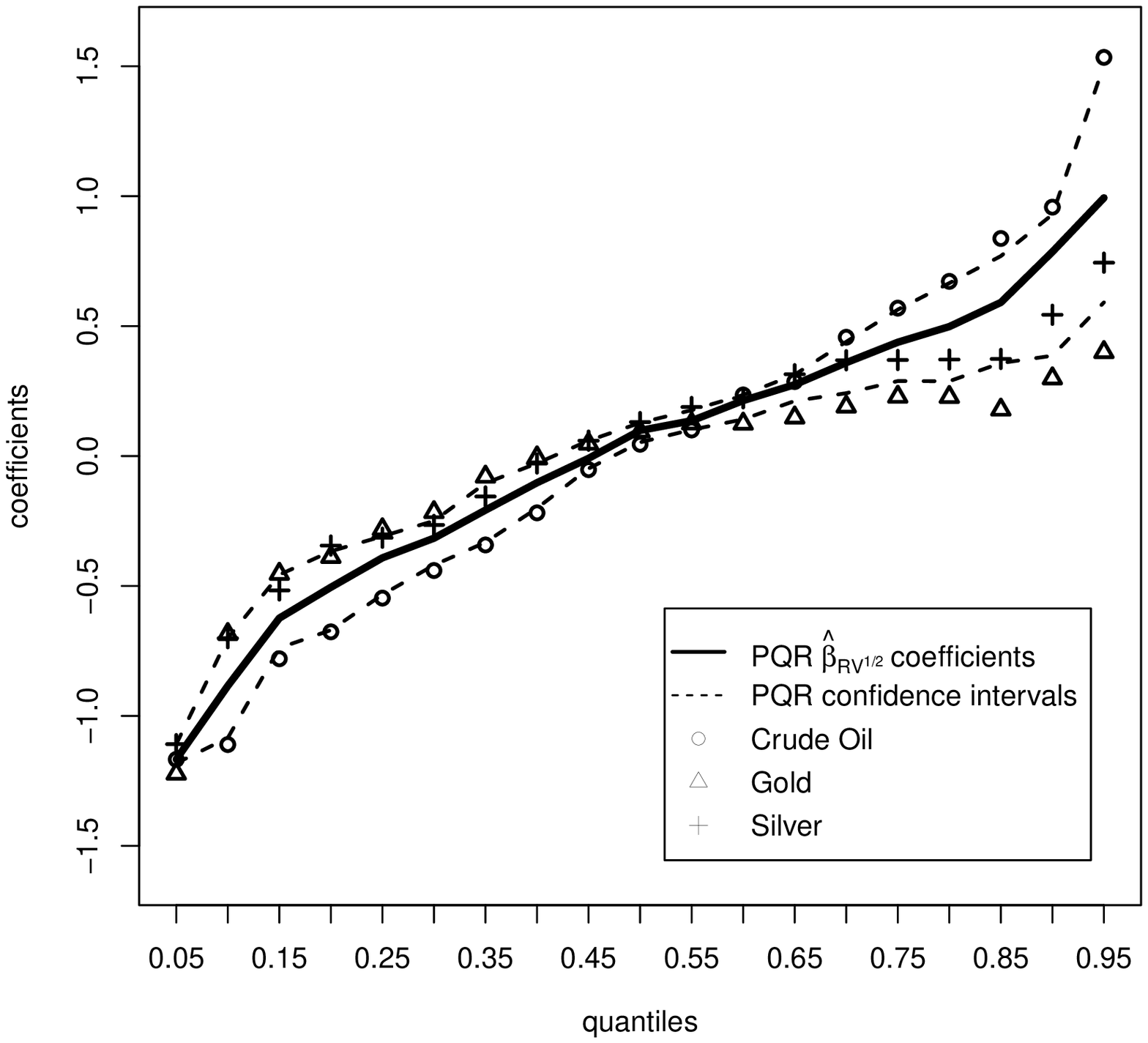}
\end{subfigure}
\begin{subfigure}[b]{0.425\textwidth}
\caption{\textit{CBOE Index}}\label{fig:PQR Index only commodity}
\vspace{-20pt}
\includegraphics[scale=0.4]{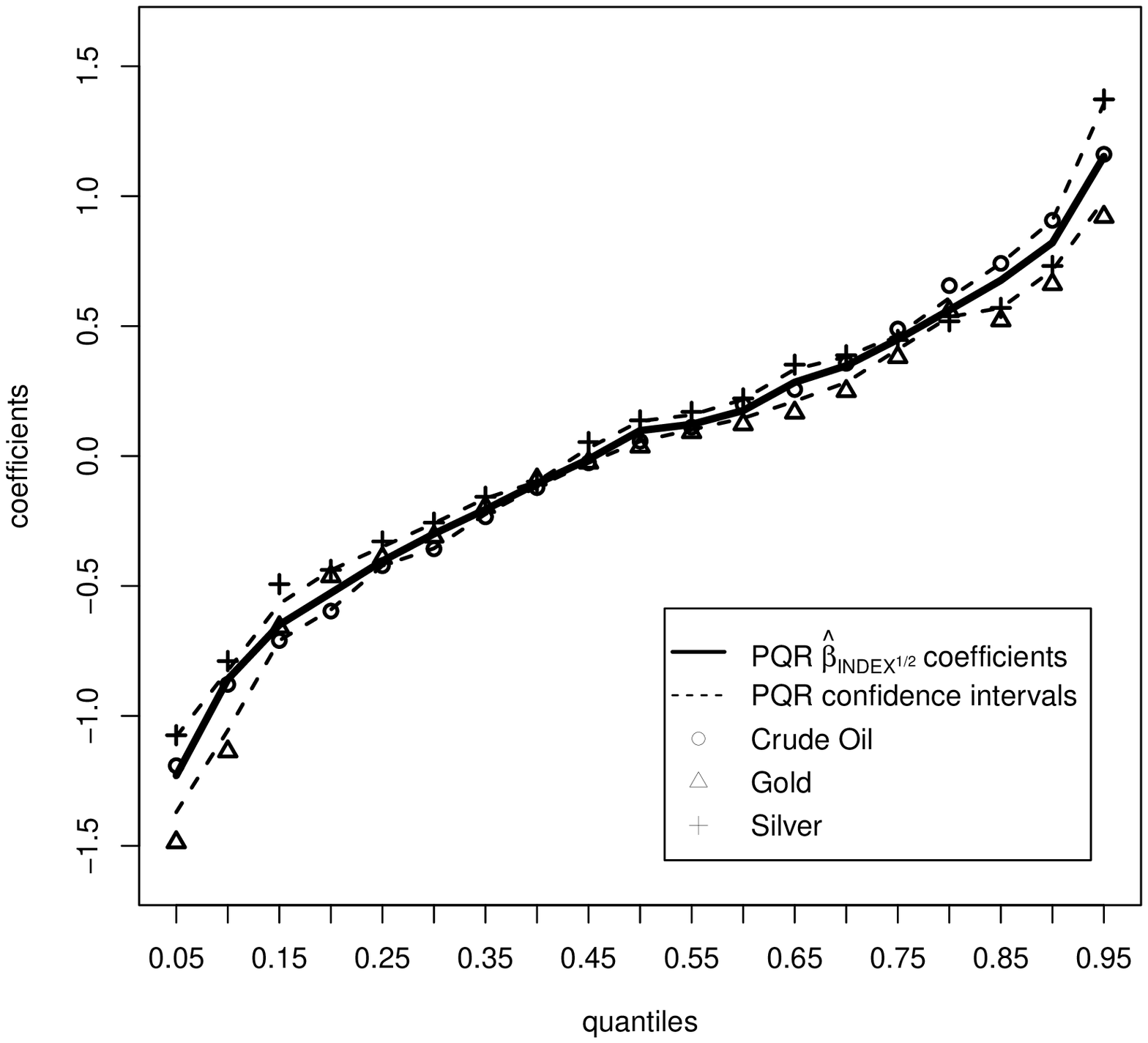}
\end{subfigure}
\captionsetup{justification=centering}
\floatfoot{Note: For both realized volatility and volatility index parameters estimates with corresponding 95\% confidence intervals are ploted by solid and dashed lines respectively. Individual univariate estimates are indicated by circle (Crude Oil), triangle (Gold), plus sign (Silver).}
\label{fig:PQR RV INDEX only}
\end{figure}

\section{Conclusion} \label{sec:conclusion}
In this paper, we propose to use Realized Volatility and CBOE Volatility Indexes together with panel quantile regression to model Value--at--Risk of the representatives of the commodity market. The choice of volatility measures allows us to directly compare difference in the impact of ex-post and ex-ante uncertainty on the conditional quantiles of commodity returns. Using panel quantile regression, we are able to control for unobserved heterogeneity among commodities and study common influence of the uncertainty measures on the Value--at--Risk estimation. The flexibility offered by panel quantile regression approach, moreover, does not require to make any distributional assumption about the distribution of commodity returns.  The advantage of our approach is revealed in the empirical application. 

In the first part of our empirical application, we study the role of the ex-post uncertainty proxied by the Realized Volatility and document common effects in the Value--at--Risk estimation of seven representatives of the commodity market. These effects hold within all studied samples, and they did not change during the Global Financial Crisis. We further show that conditional distribution of returns standardized by the lagged Realized Volatility has thinner tails than Standard Normal Distribution which is commonly used in Value--at--Risk applications.

In the second part, we complement ex-post uncertainty analysis by the option implied volatility, an ex-ante uncertainty measure. We concentrate on volatility implied by option prices since it reveals the market's sentiment and expectations about future riskiness. Due to limited availability of CBOE Volatility Indexes, we concentrate on the Crude Oil, Gold and Silver commodity alternatives to VIX index. Analogous to results from the first part, our analysis reveals similarities in patterns driving commodity Value--at--Risks. These patterns are almost identical for both ex-post and ex-ante volatility measures. In the model specification where we control for ex-post volatility, we also highlight the importance of the ex-ante uncertainty for the commodity Value--at--Risk estimation.

\bibliographystyle{chicago}
\bibliography{Econometrics}

\begin{thebibliography}{}

\bibitem[\protect\citeauthoryear{Aloui and Mabrouk}{Aloui and
  Mabrouk}{2010}]{ALOUI20102326}
Aloui, C. and S.~Mabrouk (2010).
\newblock Value-at-risk estimations of energy commodities via long-memory,
  asymmetry and fat-tailed garch models.
\newblock {\em Energy Policy\/}~{\em 38\/}(5), 2326 -- 2339.
\newblock Greater China Energy: Special Section with regular papers.

\bibitem[\protect\citeauthoryear{Andersen, Bollerslev, Diebold, and
  Labys}{Andersen et~al.}{2000}]{andersen2000exchange}
Andersen, T.~G., T.~Bollerslev, F.~X. Diebold, and P.~Labys (2000).
\newblock Exchange rate returns standardized by realized volatility are
  (nearly) gaussian.
\newblock Technical report, National Bureau of Economic Research.

\bibitem[\protect\citeauthoryear{Andersen, Bollerslev, Diebold, and
  Labys}{Andersen et~al.}{2003}]{Andersen2003}
Andersen, T.~G., T.~Bollerslev, F.~X. Diebold, and P.~Labys (2003).
\newblock Modeling and forecasting realized volatility.
\newblock {\em Econometrica\/}~{\em 71\/}(2), 579--625.

\bibitem[\protect\citeauthoryear{Ando and Bai}{Ando and
  Bai}{2017}]{ando2017quantile}
Ando, T. and J.~Bai (2017).
\newblock Quantile co-movement in financial markets; a panel quantile model
  with unobserved heterogeneity.

\bibitem[\protect\citeauthoryear{Belousova and Dorfleitner}{Belousova and
  Dorfleitner}{2012}]{belousova2012diversification}
Belousova, J. and G.~Dorfleitner (2012).
\newblock On the diversification benefits of commodities from the perspective
  of euro investors.
\newblock {\em Journal of Banking \& Finance\/}~{\em 36\/}(9), 2455--2472.

\bibitem[\protect\citeauthoryear{Bodie and Rosansky}{Bodie and
  Rosansky}{1980}]{bodie1980risk}
Bodie, Z. and V.~I. Rosansky (1980).
\newblock Risk and return in commodity futures.
\newblock {\em Financial Analysts Journal\/}~{\em 36\/}(3), 27--39.

\bibitem[\protect\citeauthoryear{Bollerslev, Hood, Huss, and
  Pedersen}{Bollerslev et~al.}{2018}]{bollerslev2016risk}
Bollerslev, T., B.~Hood, J.~Huss, and L.~H. Pedersen (2018).
\newblock Risk everywhere: Modeling and managing volatility.
\newblock {\em The Review of Financial Studies\/}~{\em 31\/}(7), 2729--2773.

\bibitem[\protect\citeauthoryear{Cabedo and Moya}{Cabedo and
  Moya}{2003}]{DAVIDCABEDO2003239}
Cabedo, J.~D. and I.~Moya (2003).
\newblock Estimating oil price `value at risk' using the historical simulation
  approach.
\newblock {\em Energy Economics\/}~{\em 25\/}(3), 239 -- 253.

\bibitem[\protect\citeauthoryear{Charles and Darn{\'e}}{Charles and
  Darn{\'e}}{2017}]{charles2017forecasting}
Charles, A. and O.~Darn{\'e} (2017).
\newblock Forecasting crude-oil market volatility: Further evidence with jumps.
\newblock {\em Energy Economics\/}~{\em 67}, 508--519.

\bibitem[\protect\citeauthoryear{Chen, Dolado, and Gonzalo}{Chen
  et~al.}{2016}]{chen2016quantile}
Chen, L., J.~J. Dolado, and J.~Gonzalo (2016).
\newblock Quantile factor models.

\bibitem[\protect\citeauthoryear{Cheng and Xiong}{Cheng and
  Xiong}{2014}]{cheng2014financialization}
Cheng, I.-H. and W.~Xiong (2014).
\newblock Financialization of commodity markets.
\newblock {\em Annu. Rev. Financ. Econ.\/}~{\em 6\/}(1), 419--441.

\bibitem[\protect\citeauthoryear{Cheong}{Cheong}{2009}]{cheong2009modeling}
Cheong, C.~W. (2009).
\newblock Modeling and forecasting crude oil markets using arch-type models.
\newblock {\em Energy policy\/}~{\em 37\/}(6), 2346--2355.

\bibitem[\protect\citeauthoryear{Chiu, Chuang, and Lai}{Chiu
  et~al.}{2010}]{CHIU2010423}
Chiu, Y.-C., I.-Y. Chuang, and J.-Y. Lai (2010).
\newblock The performance of composite forecast models of value-at-risk in the
  energy market.
\newblock {\em Energy Economics\/}~{\em 32\/}(2), 423 -- 431.

\bibitem[\protect\citeauthoryear{Chkili, Hammoudeh, and Nguyen}{Chkili
  et~al.}{2014}]{chkili2014volatility}
Chkili, W., S.~Hammoudeh, and D.~K. Nguyen (2014).
\newblock Volatility forecasting and risk management for commodity markets in
  the presence of asymmetry and long memory.
\newblock {\em Energy Economics\/}~{\em 41}, 1--18.

\bibitem[\protect\citeauthoryear{Christoffersen, Lunde, and
  Olesen}{Christoffersen et~al.}{2018}]{christoffersen2018factor}
Christoffersen, P., A.~Lunde, and K.~Olesen (2018).
\newblock Factor structure in commodity futures return and volatility.
\newblock {\em Journal of Financial and Quantitative Analysis\/}.
\newblock forthcoming.

\bibitem[\protect\citeauthoryear{Conover, Jensen, Johnson, and Mercer}{Conover
  et~al.}{2010}]{conover2010now}
Conover, C.~M., G.~R. Jensen, R.~R. Johnson, and J.~M. Mercer (2010).
\newblock Is now the time to add commodities to your portfolio?
\newblock {\em The Journal of Investing\/}~{\em 19\/}(3), 10--19.

\bibitem[\protect\citeauthoryear{Cont}{Cont}{2001}]{cont2001empirical}
Cont, R. (2001).
\newblock Empirical properties of asset returns: stylized facts and statistical
  issues.
\newblock {\em Quantitative Finance\/}~{\em 1\/}(2), 223--236.

\bibitem[\protect\citeauthoryear{Engle and Manganelli}{Engle and
  Manganelli}{2004}]{engle2004caviar}
Engle, R.~F. and S.~Manganelli (2004).
\newblock Caviar: Conditional autoregressive value at risk by regression
  quantiles.
\newblock {\em Journal of Business \& Economic Statistics\/}~{\em 22\/}(4),
  367--381.

\bibitem[\protect\citeauthoryear{Fama}{Fama}{1970}]{fama1970efficient}
Fama, E.~F. (1970).
\newblock Efficient capital markets: A review of theory and empirical work.
\newblock {\em The journal of Finance\/}~{\em 25\/}(2), 383--417.

\bibitem[\protect\citeauthoryear{Giot and Laurent}{Giot and
  Laurent}{2003}]{giot2003market}
Giot, P. and S.~Laurent (2003).
\newblock Market risk in commodity markets: a var approach.
\newblock {\em Energy Economics\/}~{\em 25\/}(5), 435--457.

\bibitem[\protect\citeauthoryear{Giot and Laurent}{Giot and
  Laurent}{2004}]{GIOT2004379}
Giot, P. and S.~Laurent (2004).
\newblock Modelling daily value-at-risk using realized volatility and arch type
  models.
\newblock {\em Journal of Empirical Finance\/}~{\em 11\/}(3), 379 -- 398.

\bibitem[\protect\citeauthoryear{Giot and Laurent}{Giot and
  Laurent}{2007}]{giot2007information}
Giot, P. and S.~Laurent (2007).
\newblock The information content of implied volatility in light of the
  jump/continuous decomposition of realized volatility.
\newblock {\em Journal of Futures Markets: Futures, Options, and Other
  Derivative Products\/}~{\em 27\/}(4), 337--359.

\bibitem[\protect\citeauthoryear{Gorton and Rouwenhorst}{Gorton and
  Rouwenhorst}{2006}]{gorton2006facts}
Gorton, G. and K.~G. Rouwenhorst (2006).
\newblock Facts and fantasies about commodity futures.
\newblock {\em Financial Analysts Journal\/}~{\em 62\/}(2), 47--68.

\bibitem[\protect\citeauthoryear{Hung, Lee, and Liu}{Hung
  et~al.}{2008}]{HUNG20081173}
Hung, J.-C., M.-C. Lee, and H.-C. Liu (2008).
\newblock Estimation of value-at-risk for energy commodities via fat-tailed
  garch models.
\newblock {\em Energy Economics\/}~{\em 30\/}(3), 1173 -- 1191.

\bibitem[\protect\citeauthoryear{Jorion}{Jorion}{2007}]{jorion2007value}
Jorion, P. (2007).
\newblock {\em Value at risk: the new benchmark for controlling market risk 3rd
  ed.}
\newblock McGraw Hill Professional.

\bibitem[\protect\citeauthoryear{Koenker}{Koenker}{2004}]{koenker2004quantile}
Koenker, R. (2004).
\newblock Quantile regression for longitudinal data.
\newblock {\em Journal of Multivariate Analysis\/}~{\em 91\/}(1), 74--89.

\bibitem[\protect\citeauthoryear{Koenker and Bassett~Jr}{Koenker and
  Bassett~Jr}{1978}]{koenker1978regression}
Koenker, R. and G.~Bassett~Jr (1978).
\newblock Regression quantiles.
\newblock {\em Econometrica: journal of the Econometric Society\/}, 33--50.

\bibitem[\protect\citeauthoryear{Li, Hurn, and Clements}{Li
  et~al.}{2017}]{LI201760}
Li, Z., A.~Hurn, and A.~Clements (2017).
\newblock Forecasting quantiles of day-ahead electricity load.
\newblock {\em Energy Economics\/}~{\em 67}, 60 -- 71.

\bibitem[\protect\citeauthoryear{Liu, Patton, and Sheppard}{Liu
  et~al.}{2015}]{liu2015does}
Liu, L.~Y., A.~J. Patton, and K.~Sheppard (2015).
\newblock Does anything beat 5-minute rv? a comparison of realized measures
  across multiple asset classes.
\newblock {\em Journal of Econometrics\/}~{\em 187\/}(1), 293--311.

\bibitem[\protect\citeauthoryear{Longerstaey and Spencer}{Longerstaey and
  Spencer}{1996}]{Longerstaey1996}
Longerstaey, J. and M.~Spencer (1996).
\newblock {RiskMetricsTM} - technical document.
\newblock {\em Morgan Guaranty Trust Company of New York: New York\/}.

\bibitem[\protect\citeauthoryear{Lux, Segnon, and Gupta}{Lux
  et~al.}{2016}]{LUX2016117}
Lux, T., M.~Segnon, and R.~Gupta (2016).
\newblock Forecasting crude oil price volatility and value-at-risk: Evidence
  from historical and recent data.
\newblock {\em Energy Economics\/}~{\em 56}, 117 -- 133.

\bibitem[\protect\citeauthoryear{Marimoutou, Raggad, and Trabelsi}{Marimoutou
  et~al.}{2009}]{marimoutou2009extreme}
Marimoutou, V., B.~Raggad, and A.~Trabelsi (2009).
\newblock Extreme value theory and value at risk: application to oil market.
\newblock {\em Energy Economics\/}~{\em 31\/}(4), 519--530.

\bibitem[\protect\citeauthoryear{Reboredo and Ugolini}{Reboredo and
  Ugolini}{2016}]{REBOREDO201633}
Reboredo, J.~C. and A.~Ugolini (2016).
\newblock Quantile dependence of oil price movements and stock returns.
\newblock {\em Energy Economics\/}~{\em 54}, 33 -- 49.

\bibitem[\protect\citeauthoryear{\v{C}ech and Barun\'{\i}k}{\v{C}ech and
  Barun\'{\i}k}{2017}]{cech2017measurement}
\v{C}ech, F. and J.~Barun\'{\i}k (2017).
\newblock Measurement of common risk factors: A panel quantile regression model
  for returns.
\newblock {\em IES WP 20/2017\/}.
\newblock Working paper, available at
  \url{http://ies.fsv.cuni.cz/default/file/download/id/31395}.

\bibitem[\protect\citeauthoryear{White, Kim, and Manganelli}{White
  et~al.}{2015}]{white2015var}
White, H., T.-H. Kim, and S.~Manganelli (2015).
\newblock Var for var: Measuring tail dependence using multivariate regression
  quantiles.
\newblock {\em Journal of Econometrics\/}~{\em 187\/}(1), 169--188.

\bibitem[\protect\citeauthoryear{Youssef, Belkacem, and Mokni}{Youssef
  et~al.}{2015}]{youssef2015value}
Youssef, M., L.~Belkacem, and K.~Mokni (2015).
\newblock Value-at-risk estimation of energy commodities: A long-memory
  garch--evt approach.
\newblock {\em Energy Economics\/}~{\em 51}, 99--110.

\bibitem[\protect\citeauthoryear{{\v{Z}}ike{\v{s}} and
  Barun{\'\i}k}{{\v{Z}}ike{\v{s}} and
  Barun{\'\i}k}{2016}]{Zikes_Barunik2015semi}
{\v{Z}}ike{\v{s}}, F. and J.~Barun{\'\i}k (2016).
\newblock Semi-parametric conditional quantile models for financial returns and
  realized volatility.
\newblock {\em Journal of Financial Econometrics\/}~{\em 14\/}(1), 185--226.

\end{thebibliography}
\pagebreak

\section*{Appendix} \label{appendix}
\begin{table}[H]
\begin{center}
\footnotesize
\caption{Coefficient estimates of Panel Quantile Regressions - Realized Volatility } \label{tab:insample_PQR}
\begin{tabular}{lccccccc}
\toprule
$\tau$ & 5\% & 10\% & 25\% & 50\% & 75\% & 90\% & 95\% \\ \cmidrule{2-8} \\[-0.75em]
 & \multicolumn{7}{c}{\textit{crisis: 2006-2010}} \\ \cmidrule{2-8}
$\hat{\beta}_{RV^{1/2}}$&-0.986 & -0.831 & -0.402 & 0 & 0.339 & 0.8 & 0.975 \\ 
 & (-9.13) & (-8.39) & (-6.07) & (0) & (6.13) & (10.18) & (6.91) \\ 
$\hat{\alpha}_{CL}$ & -0.012 & -0.009 & -0.004 & 0.001 & 0.006 & 0.009 & 0.011 \\ 
 & (-5.86) & (-4.56) & (-3.21) & (1.53) & (6.88) & (6.34) & (4.78) \\ 
$\hat{\alpha}_{CN}$ & -0.013 & -0.007 & -0.003 & 0 & 0.005 & 0.009 & 0.012 \\ 
 & (-9.62) & (-5.86) & (-2.77) & (0) & (6.02) & (7.7) & (5.59) \\ 
$\hat{\alpha}_{CT}$ & -0.012 & -0.007 & -0.004 & 0 & 0.004 & 0.007 & 0.011 \\ 
 & (-9.15) & (-5.07) & (-3.72) & (0) & (5.26) & (5.73) & (4.64) \\ 
$\hat{\alpha}_{GC}$ & -0.006 & -0.003 & -0.001 & 0.001 & 0.002 & 0.003 & 0.005 \\ 
 & (-6.33) & (-3.84) & (-2.38) & (1.61) & (5.81) & (5.3) & (4.42) \\ 
$\hat{\alpha}_{HG}$ & -0.01 & -0.006 & -0.002 & 0.001 & 0.004 & 0.005 & 0.008 \\ 
 & (-6.69) & (-5.16) & (-1.97) & (2.6) & (5.63) & (5.37) & (4.3) \\ 
$\hat{\alpha}_{NG}$ & -0.02 & -0.015 & -0.009 & -0.001 & 0.008 & 0.011 & 0.018 \\ 
 & (-7.53) & (-6.92) & (-6.43) & (-1.36) & (5.75) & (5.6) & (5.49) \\ 
$\hat{\alpha}_{SV}$ & -0.012 & -0.006 & -0.003 & 0.001 & 0.005 & 0.007 & 0.01 \\ 
 & (-6.48) & (-4.02) & (-2.48) & (2.27) & (6.69) & (5.12) & (5.46) \\ \cmidrule{2-8} \\[-0.75em]

 & \multicolumn{7}{c}{\textit{after crisis: 2011-2015}} \\ \cmidrule{2-8} 
$\hat{\beta}_{RV^{1/2}}$ & -0.949 & -0.709 & -0.304 & 0.036 & 0.292 & 0.613 & 0.878 \\ 
 & (-11.05) & (-9.17) & (-4.38) & (1.23) & (4.99) & (5.12) & (4.91) \\ 
$\hat{\alpha}_{CL}$ & -0.011 & -0.008 & -0.004 & 0 & 0.003 & 0.007 & 0.009 \\ 
 & (-11.11) & (-6.73) & (-5.22) & (0.11) & (5.32) & (5.11) & (4.04) \\ 
$\hat{\alpha}_{CN}$ & -0.009 & -0.006 & -0.004 & 0 & 0.004 & 0.007 & 0.009 \\ 
 & (-11.44) & (-7.11) & (-5.76) & (-1.19) & (6.26) & (5.92) & (5.06) \\ 
$\hat{\alpha}_{CT}$ &  -0.009 & -0.006 & -0.003 & -0.001 & 0.002 & 0.005 & 0.007 \\ 
 & (-10.41) & (-7.73) & (-4.15) & (-3.34) & (3.51) & (4.52) & (4.09) \\ 
$\hat{\alpha}_{GC}$ & -0.006 & -0.004 & -0.002 & -0.001 & 0.001 & 0.004 & 0.005 \\ 
 & (-11.52) & (-7.45) & (-4.05) & (-3.75) & (3.7) & (5.54) & (5) \\ 
$\hat{\alpha}_{HG}$ & -0.005 & -0.004 & -0.002 & 0 & 0.002 & 0.005 & 0.006 \\ 
 & (-9.98) & (-7.37) & (-4.31) & (-1.07) & (4.93) & (4.83) & (4.84) \\ 
$\hat{\alpha}_{NG}$ & -0.015 & -0.011 & -0.007 & -0.002 & 0.006 & 0.012 & 0.016 \\ 
 & (-8.32) & (-9.52) & (-6.61) & (-3.69) & (6.28) & (6.27) & (5.11) \\ 
$\hat{\alpha}_{SV}$ & -0.009 & -0.006 & -0.003 & -0.001 & 0.002 & 0.007 & 0.01 \\ 
 & (-10.04) & (-7.04) & (-4.09) & (-3.36) & (3.73) & (5.08) & (5.66) \\   \cmidrule{2-8} \\[-0.75em]

 & \multicolumn{7}{c}{\textit{full sample: 2006-2015}} \\ \cmidrule{2-8} 
$\hat{\beta}_{RV^{1/2}}$ & -1.11 & -0.879 & -0.395 & 0.041 & 0.403 & 0.825 & 1.052 \\ 
 & (-15.61) & (-11.43) & (-8.32) & (1.76) & (7.1) & (12.26) & (8.11) \\ 
$\hat{\alpha}_{CL}$ & -0.01 & -0.006 & -0.003 & 0 & 0.003 & 0.006 & 0.008 \\ 
 & (-10.63) & (-5.01) & (-5.02) & (0.45) & (4.46) & (6.23) & (4.32) \\ 
$\hat{\alpha}_{CN}$ & -0.008 & -0.005 & -0.003 & 0 & 0.003 & 0.006 & 0.009 \\ 
 & (-11.71) & (-5.94) & (-5.53) & (-1.55) & (4.86) & (8.29) & (5.52) \\  
$\hat{\alpha}_{CT}$ & -0.008 & -0.005 & -0.003 & -0.001 & 0.002 & 0.005 & 0.008 \\ 
 & (-10.62) & (-5.21) & (-4.52) & (-2.65) & (2.6) & (5.31) & (4.87) \\  
$\hat{\alpha}_{GC}$ & -0.005 & -0.003 & -0.001 & 0 & 0.001 & 0.003 & 0.004 \\ 
 & (-9.88) & (-4.93) & (-3.93) & (-1.35) & (2.69) & (6.28) & (4.81) \\   
$\hat{\alpha}_{HG}$ & -0.006 & -0.003 & -0.001 & 0 & 0.002 & 0.004 & 0.005 \\ 
 & (-8.13) & (-5.11) & (-3.64) & (0.65) & (3.73) & (5.77) & (4.93) \\  
$\hat{\alpha}_{NG}$ & -0.014 & -0.011 & -0.007 & -0.002 & 0.005 & 0.01 & 0.014 \\ 
 & (-10.28) & (-6.91) & (-7.52) & (-4.4) & (4.48) & (7.31) & (5.71) \\ 
$\hat{\alpha}_{SV}$ & -0.008 & -0.005 & -0.002 & 0 & 0.003 & 0.005 & 0.009 \\ 
 & (-9.03) & (-4.79) & (-3.77) & (-1.53) & (3.23) & (6.53) & (5.93) \\   
\bottomrule
\end{tabular}
\begin{tablenotes}
\centering
\footnotesize
\item{Note: Table displays coefficient estimates with bootstraped t-statistics in parentheses.}
\end{tablenotes}
\end{center}
\end{table}

\end{document}